\documentclass[letterpaper,10pt,journal,twoside]{IEEEtran}
\IEEEoverridecommandlockouts
\usepackage{amsmath,amssymb,amsfonts}
\usepackage{algorithmic}
\usepackage[linesnumbered,ruled]{algorithm2e}
\usepackage{graphicx}
\usepackage{textcomp}
\usepackage{xcolor}
\usepackage{threeparttable}
\usepackage{tabularx}
\usepackage{booktabs}
\usepackage{multirow}
\usepackage{colortbl}
\usepackage{enumitem}
\usepackage{subfigure}
\usepackage{pifont}
\usepackage{comment}
\usepackage{array} 

\usepackage{bigstrut,multirow,rotating}
\usepackage[numbers,sort&compress]{natbib}
\usepackage{tikz}
\usepackage{threeparttable}
\usepackage[colorlinks, linkcolor=magenta, anchorcolor=green, citecolor=blue]{hyperref}

\newcommand{\hlblue}{\textcolor{black}}

\usepackage[normalem]{ulem}

\def\BibTeX{{\rm B\kern-.05em{\sc i\kern-.025em b}\kern-.08em
    T\kern-.1667em\lower.7ex\hbox{E}\kern-.125emX}}
    
\newcommand*{\circled}[1]{\lower.7ex\hbox{\tikz\draw (0pt, 0pt)%
    circle (.5em) node {\makebox[1em][c]{\small #1}};}}

\begin{document}


\title{\huge
DDC-PIM: Efficient Algorithm/Architecture Co-design for \underline{D}oubling \underline{D}ata \underline{C}apacity of SRAM-based \underline{P}rocessing-\underline{I}n-\underline{M}emory
}

\author{Cenlin~Duan,
        Jianlei~Yang,~\IEEEmembership{Senior Member,~IEEE,}
        Xiaolin~He,
        Yingjie~Qi,
        Yikun~Wang,
        Yiou~Wang,
        Ziyan~He,
        Bonan~Yan,~\IEEEmembership{Member,~IEEE,}
        Xueyan~Wang,~\IEEEmembership{Member,~IEEE,}
        Xiaotao~Jia,~\IEEEmembership{Member,~IEEE,}
        Weitao~Pan,~\IEEEmembership{Member,~IEEE,}
        and~Weisheng~Zhao,~\IEEEmembership{Fellow,~IEEE}
\thanks{Manuscript received on April 10, 2023, revised on August 22 and October 28, 2023, accepted on October 31, 2023. This work was supported in part by the National Natural Science Foundation of China (Grant No. 62072019, 62006011 and 62004011), the Fundamental Research Funds for the Central Universities and the 111 Talent Program B16001. \textit{Corresponding authors are Jianlei Yang and Weisheng Zhao.}}
\thanks{C. Duan, X. Wang, X. Jia and W. Zhao are with BDBC, Fert Beijing Research Institute, School of Integrated Circuit Science and Engineering, Beihang University, Beijing, 100191, China. E-mail: \url{weisheng.zhao@buaa.edu.cn}}
\thanks{J. Yang, X. He, Y. Qi, Yikun Wang and Yiou Wang are with BDBC, Fert Beijing Research Institute, School of Computer Science and Engineering, Beihang University, Beijing, 100191, China. E-mail: \url{jianlei@buaa.edu.cn}}
\thanks{Z. He and W. Pan are with School of Telecommunications Engineering, Xidian University, Xi'an, 710071, China.}
\thanks{B. Yan is with Institute of Artificial Intelligence, Peking University, Beijing, 100871, China.}
}

\maketitle

\begin{abstract}
Processing-in-memory (PIM), as a novel computing paradigm, provides significant performance benefits from the aspect of effective data movement reduction. 
SRAM-based PIM has been demonstrated as one of the most promising candidates due to its endurance and compatibility. 
\hlblue{However, the integration density of SRAM-based PIM is much lower than other non-volatile memory-based ones, due to its inherent 6T structure for storing a single bit.
Within comparable area constraints, SRAM-based PIM exhibits notably lower capacity.}  
Thus, aiming to unleash its capacity potential, we propose DDC-PIM, an efficient algorithm/architecture co-design methodology that effectively \underline{d}oubles the equivalent \underline{d}ata \underline{c}apacity.
At the algorithmic level, we propose a filter-wise complementary correlation (FCC) algorithm
to obtain a bitwise complementary pair.
At the architecture level, we exploit the intrinsic cross-coupled structure of 6T SRAM to store the bitwise complementary pair in their complementary states ($Q/\overline{Q}$), thereby maximizing the data capacity of each SRAM cell.
The dual-broadcast input structure and reconfigurable unit support both depthwise and pointwise convolution, adhering to the requirements of various neural networks.
\hlblue{Evaluation results show that DDC-PIM yields about $2.84\times$ speedup on MobileNetV2 and $2.69\times$ on EfficientNet-B0 with negligible accuracy loss compared with PIM baseline implementation. Compared with state-of-the-art SRAM-based PIM macros, DDC-PIM achieves up to $8.41\times$ and $2.75\times$ improvement in weight density and area efficiency, respectively.}

\end{abstract}





\IEEEpeerreviewmaketitle

\section{Introduction}\label{sec:Introduction}

\IEEEPARstart{D}{eep} Neural Networks (DNNs) are ubiquitous in a variety of applications, such as image recognition \cite{krizhevsky2017imagenet,szegedy2017inception,he2016deep}, speech recognition \cite{burchi2023audio,zhang2017very}, and object detection \cite{wang2022internimage,fang2022eva,su2022towards}. 
To achieve higher accuracy, an intuitive approach is to design deeper and more intricate network models.
However, the complexity of these models makes them difficult to deploy on edge devices, such as wearable devices, medical equipment, and mobile devices.
To tailor for mobile and resource-constrained environments, there is a growing interest in building small and efficient neural networks (NNs), such as MobileNet \cite{howard2017mobilenets} and EfficientNet-B0 \cite{tan2019efficientnet}. 
These compact NNs usually decompose standard convolution into separable convolution, significantly reducing the number of parameters and computations with negligible accuracy loss.

\begin{figure}[t]
\centering
\includegraphics[width=1.0\linewidth]{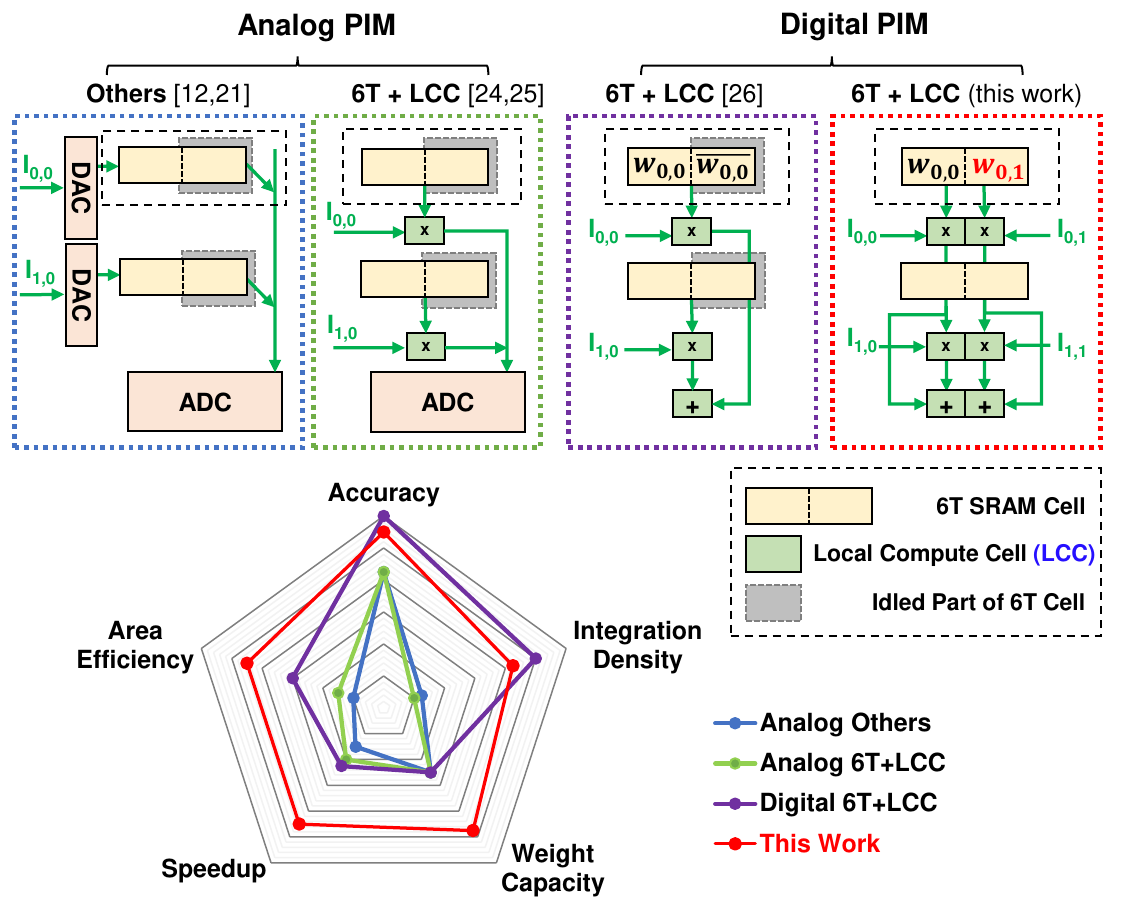}
\caption{Typical principles demonstration and comparison among several state-of-the-art 6T SRAM-based PIM studies.}
\label{fig1:Comparison}
\end{figure} 


\hlblue{While compact NNs offer the potential for enhanced efficiency, the inherent data-intensive nature of such networks continues to present challenges for edge devices. 
This is particularly evident in edge devices built on conventional Von Neumann architectures, as the deployment on such devices will bring in large energy consumption, due to frequent data movements between memory and computing units. With the saturation of Moore’s law, reducing energy consumption becomes progressively more difficult.}
As a new architectural paradigm, Processing-in-memory (PIM) performs Multiply-Accumulate (MAC) operations in memory to eliminate the \textit{Memory Wall} bottleneck, making it a promising candidate architecture for energy-efficient edge devices.
As illustrated in Fig.~\ref{fig1:Comparison}, PIM macros are usually categorized into analog PIM and digital PIM \cite{yin2021pimca,gonugondla201842pj,su202015,yan20221}.
Analog PIM usually performs MAC operations in the voltage or current domain \cite{yin2021pimca,gonugondla201842pj}.
However, the extra ADC/DAC \cite{su202015} and process variation often lead to low area efficiency, low computation parallelism, and non-negligible accuracy loss.

To tackle the above limitations, digital PIM is further proposed by performing bitwise operations \cite{yan20221}.
Generally, digital PIM integrates logic units into a single cell array, simultaneously activating all rows to improve computing parallelism.
Previous studies have shown that various technologies, such as RRAM \cite{ Imani2019DigitalPIMDP, 8980299}, SRAM and MRAM \cite{chen2022accelerating,Zhang2021TimeDomainCI,9218660,9729451}, are available candidates for digital PIM.
Among these technologies, SRAM is widely used in academia and industry due to its faster write speed, lower write energy and compatibility with proven process technologies.
However, SRAM suffers from a low integration density compared to non-volatile memory. The conventional structure of SRAM cells requires six transistors to store one bit in two cross-coupled inverters, whereas non-volatile memory is capable of storing multiple bits with fewer transistors.
Thus, inhibited by the integration density, the capacity has become a crucial limitation for SRAM-based PIM.

\begin{figure}[t]
\centering
\includegraphics[width =1.0\linewidth]{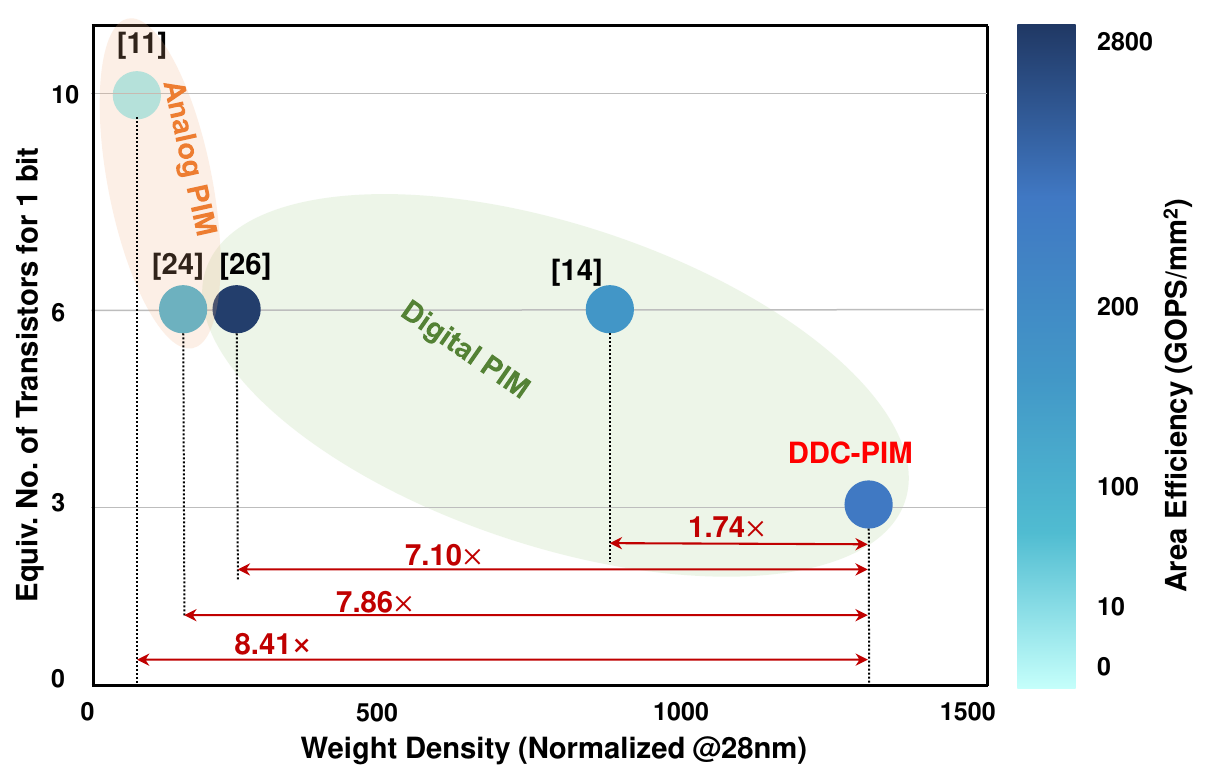}
\caption{Normalized weight density improvement and area efficiency comparison with prior SRAM-based PIM solutions.}
\label{fig2:integration}
\end{figure}

From the essential observations of adopted cross-coupled inverters in SRAM cells, a pair of complementary states (denoted as $Q$ and $\overline{Q}$) is used for storing a single bit.
In practice, only one state within the same pair will participate in the computation regarding the bit it represents.
Based on this observation, we seek to promote the representational power of complementary pairs by treating each state as an independent bit of information.
To realize this vision, the data organization must be adjusted correspondingly at the algorithmic level.
Hence, storing the weights of compact NNs in this structure should embody the above complementary characteristic.
Based on this cooperation between algorithm and architecture, the \textit{weight density} (i.e., weight capacity per area) will achieve a twofold increase.
As shown in Fig.~\ref{fig2:integration}, this approach is equivalent to using three transistors for storing one bit, which offers us an opportunity to boost the capacity of SRAM-based PIM.
As the increase in overall area for supporting the above approach is negligible, the weight density achieves remarkable improvement compared with prior works.
Moreover, compared to \cite{yan20221} with a PIM macro similar to ours, the area-efficiency of DDC-PIM also approximately doubles due to improved computation parallelism.


In this work, we propose DDC-PIM, an efficient algorithm/architecture co-design framework that comprises a novel DDC-PIM architecture and a \underline{F}ilter-wise
\underline{C}omplementary \underline{C}orrelation (FCC) algorithm. 
Essentially, DDC-PIM doubles the weight density without modifying the SRAM structure, leading to higher equivalent data capacity and area efficiency.
Our main contributions can be summarized as follows:

\begin{itemize}

\item We propose a novel DDC-PIM architecture, the first in-memory architecture for doubling the equivalent data capacity of SRAM-based PIM. By fully exploiting the intrinsic cross-coupled structure, DDC-PIM overcomes the limitation of low integration density of SRAM. 

\item We propose a FCC algorithm to obtain bitwise complementary filters. This method can reduce the memory footprint/computation latency and improve the equivalent data transfer bandwidth of NNs with negligible accuracy loss.

\item A dual-broadcast input structure and a reconfigurable unit, together with a flexible mapping strategy are co-optimized to achieve a high PIM array utilization rate.





\end{itemize}

The rest of this paper is organized as follows.
Section~\ref{sec:Background} provides background and motivations.
Section~\ref{sec:Methodology} demonstrates the details of the proposed methodology. 
Section~\ref{sec:Experiments} illustrates the experimental results.
Related discussions are illustrated in Section~\ref{sec:Discussion} and conclusion remarks are given in Section~\ref{sec:Conclusion}.
\section{Background and Motivations}
\label{sec:Background}

\subsection{Requirements from Compact NN Models}

Modern state-of-the-art NN models usually necessitate lots of memory and computational resources that are easily beyond the capabilities of typical edge devices.
Hence, there is a growing interest in developing compact and efficient NN models. 
One of the mainstream approaches is to obtain smaller models in decomposed convolution manner, minimizing the scale of parameters and computations.
\hlblue{Among them, MobileNet and EfficientNet are two typical compact NN models, which inherently possess significantly lower levels of redundancy compared to regular NNs.}
MobileNet decomposes \textit{standard convolution} (\texttt{std-conv}) into \textit{pointwise convolution} (\texttt{pw-conv}) and \textit{depthwise convolution} (\texttt{dw-conv}), which are usually collectively referred to as \textit{separable convolution} (\texttt{sp-conv}) \cite{howard2017mobilenets}. 
For \texttt{std-conv}, convolution filters are considered three-dimensional, including a channel dimension and two spatial dimensions.
For \texttt{sp-conv}, each input channel is assigned with a single filter in \texttt{dw-conv} for spatial correlation, the results of which are then combined through $1 \times 1$ convolutions in \texttt{pw-conv} for channel correlation.
Thus, the amount of computation in \texttt{sp-conv} is only about $1/8$ of \texttt{std-conv}. 
Following the idea of \texttt{sp-conv} in MobileNet, EfficientNet uniformly scales all dimensions of \textit{depth}, \textit{width} and \textit{resolution}, using highly effective compound coefficients to maximize the model efficiency \cite{tan2019efficientnet}.
EfficientNet can usually obtain a faster inference speed and better accuracy with a smaller model size than the most popular convolutional NN models. 


\subsection{Capacity Issues of SRAM-based PIM}

For earlier SRAM-based PIM macros \cite{kang2018multi,gonugondla201842pj}, as illustrated in Fig.~\ref{fig1:Comparison} (\textit{denoted by} \textbf{Analog Others}), input features are first converted into analog signals through digital to analog converters (DACs), and then taking part in analog MAC operations with the pre-stored weights in SRAM cells.
The results are then converted back to the digital domain by analog to digital converter (ADCs).
In general, multiple cell rows in PIM array would activate simultaneously to increase computational parallelism.
However, this approach will introduce \textit{read disturbance} due to the expanded voltage swing of the bitline.
One possible solution to this issue is to increase the bitline voltage, sacrificing the signal margin of read operation. 
Another approach is to use non-6T structure SRAM, such as 8T, 10T, or 12T, which also brings great area overhead \cite{sinangil20207,yin2021pimca,zhang2018recryptor}. 
Some researchers have proposed LCC-based (Local Compute Cell-based) PIM macros as shown in Fig.~\ref{fig1:Comparison} (\textit{denoted by} \textbf{Analog 6T+LCC}) to resolve the above limitations \cite{si202015,yue2022br}, where LCC means the newly integrated \textit{local computing cells}.
The performance of these PIM architectures is often limited by kinds of analog-specific non-idealities, including process variations and tremendous area/power overhead due to essential data conversions.
Moreover, handicapped by the accuracy of ADCs, only a limited number of cell rows can be activated per cycle, leading to low PIM array utilization and high computation latency. 

In contrast to analog PIM solutions which perform computations in the voltage or current domain, digital PIM
incorporates logic units within a single-cell array to execute digital logic operations. 
This approach enhances the accuracy, area efficiency, and energy efficiency of the system, as demonstrated by previous studies \cite{chih202116,tu2022trancim,jinshan2023isscc,shiweiisscc,fengbin2023isscc,fengbin2023isscctensor}.
Yu et al. \cite{chih202116} propose an all-digital SRAM-based full-precision PIM macro.
Yue et al. \cite{jinshan2023isscc} propose a floating-point CIM processor for NN inference/training applications. Yan et al. \cite{yan20221} propose an ADC-less SRAM-based PIM with reconfigurable bitwise operations as shown in Fig.~\ref{fig1:Comparison} (\textit{denoted by} \textbf{Digital 6T+LCC}).
However, only half of the 6T bitcells within these PIM macros contribute to the MAC computations, due to that a single bit is represented by two complementary states in the cross-coupled structure.
As shown in the upper right of Fig.~\ref{fig1:Comparison} (\textit{denoted by} \textbf{this work}), our intuitive approach is to treat each state as an independent bit of information to fully utilize these bitcells for \textit{doubling the equivalent data capacities}.
When a pair of computation units are connected with the cross-coupled structure, two independent \texttt{AND} operations can also be performed simultaneously, providing opportunities for \textit{doubling computation parallelism}.
As a result illustrated in the radar plot from Fig.~\ref{fig1:Comparison}, our work achieves remarkable enhancements in area efficiency, weight density and speed up, with minor trade-offs in integrated density and accuracy.

\section{Methodologies}\label{sec:Methodology}

This section presents the proposed DDC-PIM in detail, including the FCC algorithm, DDC-PIM architecture, and data mapping method, 
which seeks to harness the capacity potential and improve the area efficiency of digital 6T SRAM PIM.

\subsection{Overall Framework}\label{sec:overall}

\begin{figure}[t]
\centering
\includegraphics[width =0.8\linewidth]{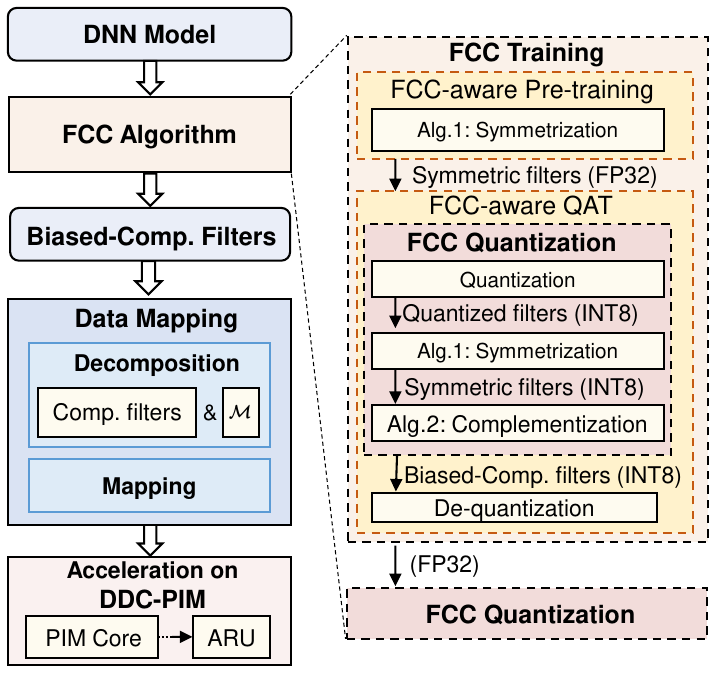}
\caption{\hlblue{Overview of proposed DDC-PIM framework.}}
\label{fig3:overview}
\end{figure}

Fig.~\ref{fig3:overview} illustrates the overview of the DDC-PIM framework, consisting of three major components: FCC algorithm, DDC-PIM architecture, and data mapping method.
\hlblue{For compact NNs, DDC-PIM first transforms the weights into a biased bitwise complementary format by the two-stage FCC algorithm, comprising FCC training and FCC quantization.
During data mapping, the results of FCC algorithm are decomposed into bitwise complementary filters and corresponding mean values layer-by-layer.
The per-layer configuration signals are also generated offline during this procedure. 
Due to the complementary nature in these filters, only half of the bitwise complementary filters are required for storage and transmission.}
DDC-PIM then performs MAC operations with these filters on the PIM core.
The convolution results are recovered from the MAC outputs and the mean value using an accumulate and recover unit (ARU). 
This framework can double data capacity and reduce computation latency by exploiting the bitwise complementary nature of weights. 
Meanwhile, only half of the complementary filters are required during data transmission, equivalently increasing the data transfer bandwidth.
Subsequently, we will detail the three main components of DDC-PIM.
 
\subsection{FCC Algorithm}\label{sec:alg}

\begin{figure*}[t]
\centering
\includegraphics[width =1.0\linewidth]{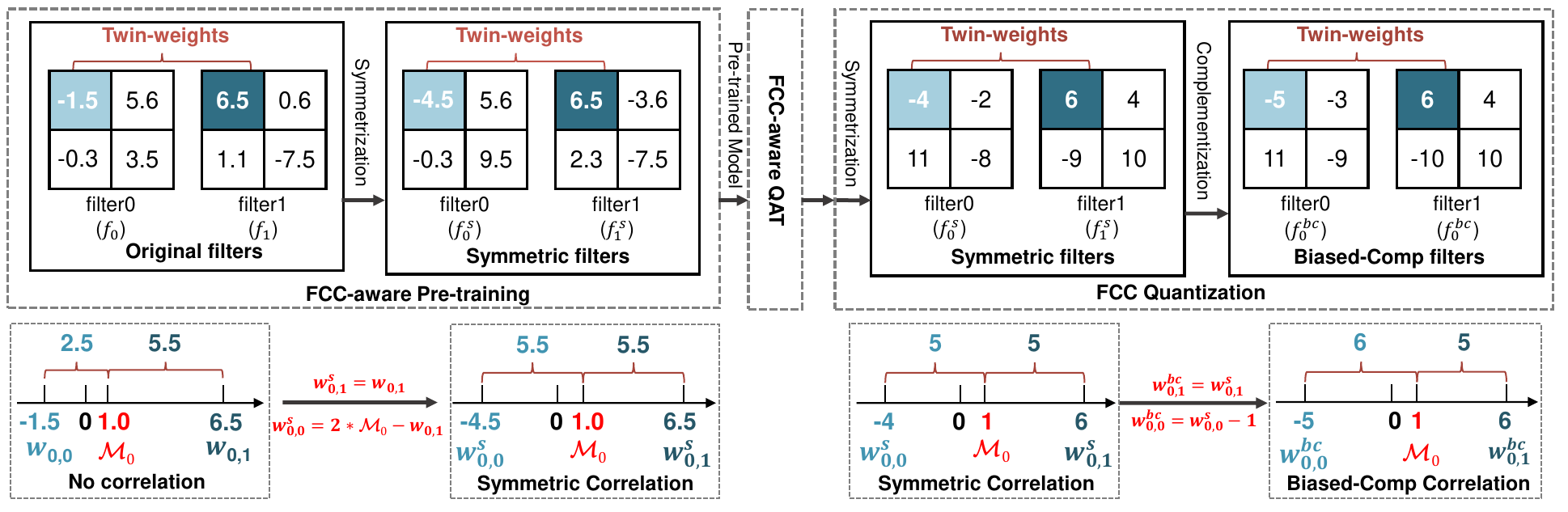}
\caption{\hlblue{Demonstration of FCC algorithm.}}
\label{fig4:Demonstrations}
\end{figure*} 

For simplicity, we first introduce the notations that will facilitate the subsequent exposition, as illustrated in Tab. \ref{tab:Meaning}.

\begin{table}[t]
\caption{Notations of the involved mathematical symbols and corresponding descriptions.}
\label{tab:Meaning}
\centering
\begin{tabular}{|p{1cm}<{\centering}|p{6cm}|}
\hline
\textbf{Term} & \textbf{Description} \\ \hline
$\mathcal{I}$     &     input feature map  \\ \hline
$\mathcal{O}$     &     output feature map   \\ \hline
$\cal{F} $ & original filters \\ \hline
${\cal{F}}^s $ & symmetric filters \\ \hline
${\cal{F}}^{bc} $ & biased-comp filters \\ \hline
$f_j^{(s)} $ &  $j^{th}$ filter with symmetric correlation  \\ \hline
$f_j^{(c)} $ & $j^{th}$ filter with complementary correlation     \\ \hline
$f_j^{(bc)} $ & $j^{th}$ filter with bias-complementary correlation \\ \hline
$w_{i,j}$  & $i^{th}$ weight in $j^{th}$ filter \\ \hline
$\mathcal{M}$     &   mean value of a pair of adjacent filters      \\ \hline
 $\ast$  &   convolution operation function      \\ \hline
 $\sim$  &   bitwise complement operation     \\ \hline
\end{tabular}
\end{table}

A convolutional (\texttt{Conv}) layer applies $N$ 3-dimensional (3D, $K \times K\times C$) filters ($\cal{F}$) on 3D ($H \times W\times C$) input feature maps ($\mathcal{I}$) to extract embedded characteristics and generate the output feature maps ($\mathcal{O}$).
$H$ and $W$ denote the height/width of $\mathcal{I}$. $C$ and $N$ denote the number of channels of $\mathcal{I}$ and $\mathcal{O}$, respectively.
$K$ denotes the kernel size.
As illustrated in Fig.~\ref{fig4:Demonstrations}, we use two adjacent filters ($f_0$ and $f_1$) with the size of $2 \times 2\times 1$ as an example, where ${\cal M}_0$ denotes the mean value of $f_0$ and $f_1$.
The weights located in the same position of these two filters are denoted as \textit{twin-weights}, i.e. $w_{i,j}$ and $w_{i,j+1}$, where $i\in \left [ 0, 1, 2, \ldots, K\times K\times C -1 \right ]$ and $j \in [ 0, 2, 4, \ldots,  N-2  \rfloor$.
And a filter pair comprising twin-weights with symmetric correlation is denoted as \textit{symmetric filters}. More precisely, twin-weights in \textit{symmetric filters} satisfy: 
\begin{equation}\label{eq:eq1}
    \begin{split}
         & w_{i,j}^{s} - \mathcal{M} \enspace = \enspace -(w_{i,j+1}^{s} - \mathcal{M}) .
    \end{split}
\end{equation}   
A filter pair composed of bitwise complementary twin-weights is denoted as \textit{complementary filters (Comp filters)}, which satisfies:
\begin{equation}\label{eq:eq2}
    \begin{split}
         w_{i,j}^{c} \enspace = \enspace \sim w_{i,j+1}^{c} .
    \end{split}
\end{equation}
A pair of complementary filters with a fixed bias (equals to $\cal M$) is denoted as \textit{biased-complementary filters (Biased-Comp filters)}, which satisfies:
\begin{equation}\label{eq:eq3}
    \begin{split}
         w_{i,j}^{bc} - \mathcal{M} \enspace=\enspace \sim (w_{i,j+1}^{bc} - \mathcal{M}) .
    \end{split}
\end{equation} 
By storing the complementary bits ($w^c_{i,j}[k]$ and $w^c_{i,j+1}[k]$, $k\in \left [ 0, 1, 2, \ldots, 7 \right ]$) in the cross-coupled structure of 6T SRAM, we only need to transfer half of the filters and a group of $\mathcal{M}$. 
Thus, the data transfer bandwidth can be equivalently improved by about $2\times$ with little extra overhead. 

\begin{algorithm}[t]
\KwIn{Original or quantized filters ${\cal{F}} \doteq \left[ f_0, \dots, f_{N-1} \right]$, where $C/N$ are the numbers of input/output channels, $K$ is kernel size.}
\KwOut{Symmetric filters ${\cal{F}}^s \doteq \left[ f^s_0,\dots,f^s_{N-1} \right]$.}
    ${\cal L} \gets K\times K\times C$ \\
    \For{$j$ \bf{in} $[ 0, 2, 4, \dots, N-1 \rfloor$}{
        $sum_j \gets \left( \sum f_{j} + \sum f_{j+1} \right)$ \\
        ${\cal M}_j \gets sum_j / (2 \times {\cal L}) $ \\
        \eIf {$\vert f_j(\cdot) - {\cal M}_j\vert \ge \vert f_{j+1}(\cdot) - {\cal M}_j\vert $ }
        {
          $f^s_j(\cdot) \gets f_j(\cdot)$ \\
          $f^s_{j+1}(\cdot) \gets  \left( 2 \times {\cal M}_j - f_j(\cdot) \right)$
        }{
          $f^s_j(\cdot) \gets \left( 2 \times {\cal M}_j - f_{j+1}(\cdot) \right)$ \\
          $f^s_{j+1}(\cdot) \gets f_{j+1}(\cdot)$
        }
    }
\caption{Symmetrization}
\label{alg1}
\end{algorithm}

\hlblue{FCC algorithm includes FCC training and FCC quantization, as shown in Fig.~\ref{fig3:overview}.
FCC training procedure mainly consists of two steps, FCC-aware Pre-training and FCC-aware QAT, as following.}

\subsubsection{\hlblue{FCC-aware Pre-training}}

We pre-train the NN models to acquire \textit{symmetric filters} through \textit{Symmetrization} during pre-training, as shown in Alg.~\ref{alg1}. 
Considering bitwise operations are usually incompatible with floating point numbers, FCC uses negations to approximate the bitwise complementary operations during pre-training.
As shown in Eq.~\ref{eq:eq4}, the opposite value of an integral number is similar to its bitwise complement, making it a viable solution for approximation before quantization.
\begin{equation}\label{eq:eq4}
    \begin{split}
         -w_{i,j} \enspace=\enspace \sim w_{i,j} + 1  .
    \end{split}
\end{equation}   

However, the constraint of symmetrizing all filter pairs to zero is too rigid for NN models to extract enough information during training.
Hence, it will introduce drastic accuracy loss.
For instance, such a restriction resulted in MobileNetV2 accuracy plummeted from $97.05\%$ to $71.65\%$.
To alleviate the impact of such restriction, we propose to symmetrize filter pairs to their mean value $\cal M$ for more flexibility.
Fig.~\ref{fig4:Demonstrations} demonstrates an illustrative example of our proposed FCC algorithm.
The procedure of FCC-aware pre-training mainly consists of the following two steps:

\begin{enumerate}[label=\large\protect\textcircled{\small\arabic*}]
\item First, we initialize the original filters by randomly sampling a normal distribution and then calculate $\cal M$ for each adjacent filter pair.
In this case, ${\cal M}_0 = 1.0$, $w_{0,0} = -1.5$, and $w_{0,1} = 6.5$.
\item Then, we symmetrize these twin-weights according to $\cal M$ by replacing the weight closer to $\cal M$ with the mirror image of the other, as shown in Fig.~\ref{fig4:Demonstrations}.
After symmetrization, $w^s_{0,0}$ and $w^s_{0,1}$ are represented as $-4.5$ and $6.5$, respectively.
\end{enumerate}

As a result of pre-training, the obtained twin-weights in \textit{symmetric filters} exhibit an opposite correlation when $\cal M$ is extracted, which satisfies:  
\begin{equation}\label{eq:eq5}
w^s_{0,0} - {\cal M}_0 \enspace=\enspace -(w^s_{0,1} - {\cal M}_0) .
\end{equation}   


\subsubsection{\hlblue{FCC-aware QAT}}

\hlblue{
Quantization-aware training (QAT) simulates the quantization effect by applying quantization and subsequent de-quantization during the training process.
In order to perceive the impact of complementary operations on model accuracy, we modify QAT to a FCC-aware manner for obtaining quantization parameters of the above pre-trained model.
Specifically, the procedure of FCC-aware QAT includes the following four steps:}

\begin{algorithm}[t]
\KwIn{Symmetric filters ${\cal{F}}^s \doteq \left[ f^s_0,\dots,f^s_{N-1} \right]$, where $N$ is the number of output channels.}
\KwOut{Biased-Comp filters ${\cal{F}}^{bc} \doteq \left[ f^{bc}_0,\dots,f^{bc}_{N-1} \right]$. }
\SetKwFunction{}{}
    \For{$j$ \bf{in} $[ 0, 2, 4, \dots, N-2 \rfloor$}{
        \eIf {${f^s_j}(\cdot)\ge f^s_{j+1}(\cdot)$}
        {
          $f^{bc}_{j+1}(\cdot) \gets \left( f^s_{j+1}(\cdot)-1 \right)$ \\
          $f^{bc}_{j}(\cdot) \gets f^s_{j}(\cdot)$ \\
        }{
          $f^{bc}_j(\cdot) \gets \left( f^s_j(\cdot)-1 \right)$ \\
          $f^{bc}_{j+1}(\cdot) \gets f^s_{j+1}(\cdot)$
        }
}
\caption{Complementization}
\label{alg2}
\end{algorithm}

\begin{enumerate}[label=\large\protect\textcircled{\small\arabic*}]
\item \textit{\hlblue{Quantization}}: \hlblue{We quantize a floating point model (\textit{symmetric filters}) to an 8-bit precision model (\textit{quantized filters}) for calculating the quantization parameters.}
\item \textit{Symmetrization}: On account of the weakened symmetric correlation within the \textit{quantized filters}, we perform symmetrization for a second time. When symmetrization is performed, $\cal M$ is rounded to ensure that $\cal M$ is an integer. As illustrated in Fig.~\ref{fig4:Demonstrations}, after quantization and symmetrization, $w^s_{0,0} = -4$, $w^s_{0,1} = 6$, and ${\cal M}_0 = 1$.
\item \textit{Complementization}: Due to the relationship between the opposite and the bitwise complementary value of $w_{i,j}$ presented in Eq.~\ref{eq:eq4}, we exert \textit{Complementization} on \textit{symmetric filters} (\texttt{INT8}) to obtain \textit{Biased-Comp filters}, as shown in Alg.~\ref{alg2}.
Specifically, we subtract $1$ from the smaller twin-weight in \textit{symmetric filters} to obtain the bitwise complementary correlation after $\cal M$ is extracted. After complementization, $w^{bc}_{0,0}$ and $w^{bc}_{0,1}$ become $-5$ and $6$ respectively.
\hlblue{\item \textit{De-quantization}: We perform de-quantization operation on \textit{Biased-Comp filters}, in order to ensure accurate gradient calculations and parameter updates.}
\end{enumerate}

\hlblue{We refer to the above process of quantization, symmetrization and complementization as FCC quantization.
At the end of FCC training, we perform FCC quantization again to obtain \textit{biased-comp filters.}}
The FCC algorithm leverages the complementary characteristics of weights in \texttt{Conv} layers to accommodate the DDC-PIM architecture, which offers a novel paradigm for deep learning computation. 
However, the fully connected (\texttt{FC}) layers do not benefit from this algorithm as much as the \texttt{Conv} layers do, which will be detailed in Section~\ref{sec:Experiments}. Therefore, we exclude the \texttt{FC} layers from the scope of the FCC algorithm.
After the biased-comp. filters generation module, the \textit{Biased-Comp filters} will be decomposed and mapped onto the DDC-PIM architecture, which enables parallel computation by employing the cross-coupled structures of 6T SRAM.   

\begin{figure}[t]
\centering
\includegraphics[width = 1.0\linewidth]{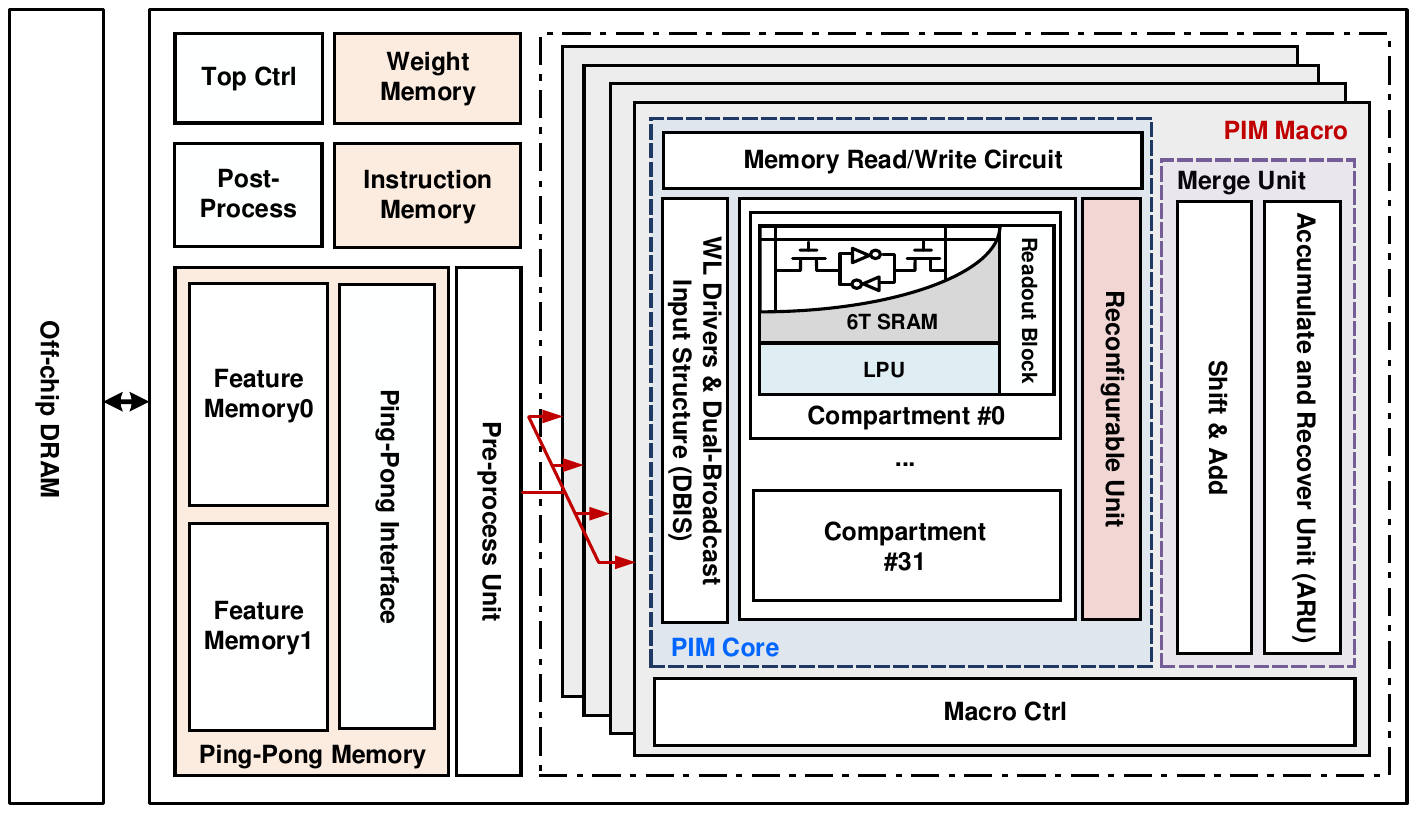}
\caption{Top level architecture design of DDC-PIM.}
\label{fig5:architecture}
\end{figure}

\begin{figure*}[t]
\centering
\includegraphics[width = 1\linewidth]{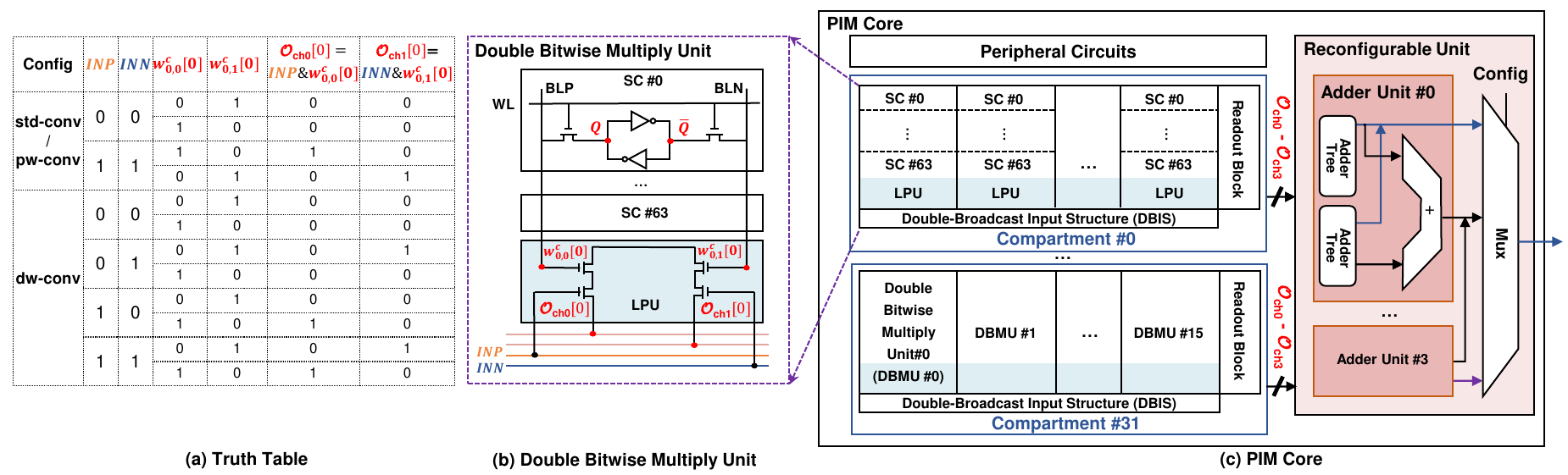}
\caption{Circuit design of PIM core and its corresponding truth table.}
\label{fig6:pim_core}
\vspace{-9pt}
\end{figure*}

\subsection{Architecture Design of DDC-PIM}\label{sec:arch}

\subsubsection{Top level architecture}\label{sec:top_arch}

Fig.~\ref{fig5:architecture} shows the overall architecture of DDC-PIM, which is composed of a top controller, a pre-process unit, four PIM macros, an instruction memory, a weight memory, a ping-pong memory, and a post-process unit. 
The PIM macro is an extension of the ADC-less SRAM PIM macro proposed in \cite{yan20221}.
The DDC-PIM is interfaced with an off-chip DRAM that stores the input features and weights of a neural network.
To accomplish the execution of a convolutional layer, the input features, weights and instructions are fetched from off-chip DRAM to on-chip memories.
After transmission, the top controller first processes instructions fetched from instruction memory and sends corresponding control signals to the whole system.
$\cal I$ stored in ping-pong memory can be accessed by pre-process unit for converting into bit-serial form and broadcasting to four PIM macros.
The PIM core in each macro receives weights from \textit{weight memory} and performs bitwise MAC operation with $\cal I$ to generate the intermediate results.
These results are shifted and accumulated by shift \& add unit based on their respective bit position for producing the \textit{partial sum} (Psum).
Then, we accumulate ($\sum {\cal I} )\times \cal M$ and Psum in ARU to recover the convolution results. 
The results are sent to post-process unit for performing pooling and other operations.
Finally, $\cal O$ acquired from the post-process unit is written back to the ping-pong memory.
Details of some key modules are presented below.

\subsubsection{PIM core with dual-broadcast input structure (DBIS) and reconfigurable unit}\label{subsubsec:Architure}

As listed in Fig.~\ref{fig6:pim_core}(c), each PIM core consists of $32$ compartments, a reconfigurable unit, and other peripheral circuits.
Each compartment comprises $16$ double-bitwise multiply units (DBMUs) for four signed \texttt{8 bit} weights configurations ($w^c_{0,0},w^c_{0,1},w^c_{0,2},w^c_{0,3}$), a DBIS supporting two different inputs ($INN$ and $INP$), and a readout block.
Each DBMU consists of sixty-four 6T SRAM cells (SC \#0 – SC \#63) and one local processing unit (LPU), as illustrated in Fig.~\ref{fig6:pim_core}(a) and Fig.~\ref{fig6:pim_core}(b). Sixty-four SCs in one column share one LPU and perform two bitwise \texttt{AND} operations of \texttt{1 bit} input and \texttt{1 bit} weight with the support of DBIS.    
The results generated by LPU are sampled and held by the readout block. The reconfigurable unit and the DBIS endow the PIM core with the flexibility of executing various convolution operations.
Each PIM core can be operated in three operation modes:
\raisebox{.5pt}{\textcircled{\raisebox{-.9pt} {1}}} normal SRAM mode supporting either read or write operations, \raisebox{.5pt}{\textcircled{\raisebox{-.9pt} {2}}} regular computing mode for multi-bit MAC operations, and \raisebox{.5pt}{\textcircled{\raisebox{-.9pt} {3}}} double computing mode with dual-broadcast inputs for multi-bit MAC operations.

\begin{figure}[t]
\centering
\includegraphics[width = 1.0\linewidth]{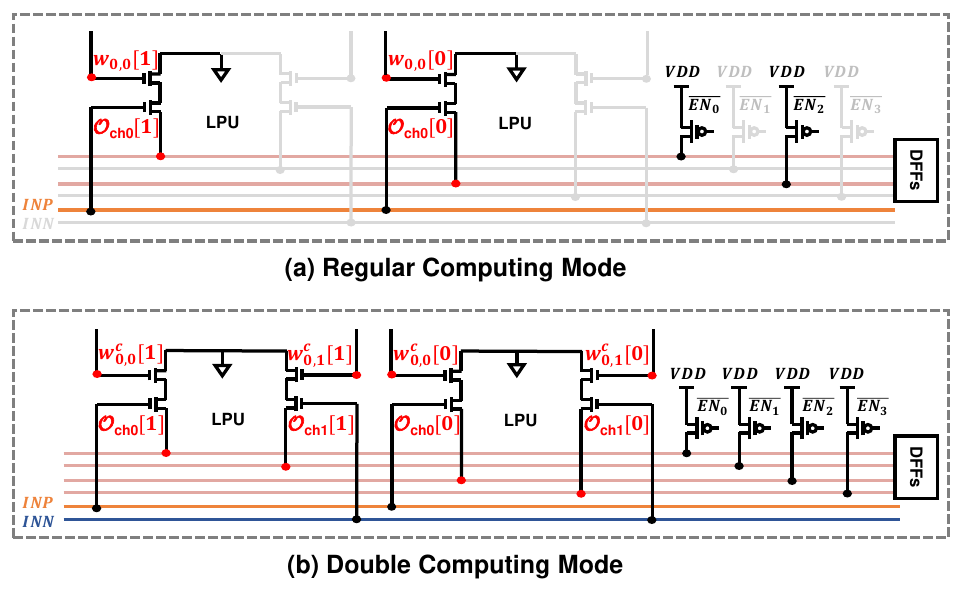}
\caption{Illustration of computation modes for PIM core.}
\label{fig8:PIMMode}
\end{figure}

In normal SRAM mode, the corresponding bits stored in $Q$ and $\overline{Q}$ can be accessed through BL pairs (BLP/BLN) in each standard 6T SRAM bitcell for read and write operation.  

In regular computing mode, only dynamic logic switches $\overline{EN_0}$ and $\overline{EN_2}$ are set to ground for pre-charging output to $VDD$ during each computing cycle.
As illustrated in Fig.~\ref{fig8:PIMMode}(a), the black lines represent an available path and the gray lines represent an inaccessible path, indicating only half of the LPU enable and perform multiplication of binary inputs and weights.
A corresponding DFF samples and holds the computational result, which satisfies:
\begin{equation}
\label{eq:eq6}
O_{ch0}[1] \enspace=\enspace w_{0,0}[1] \enspace \& \enspace INP  .
\end{equation} 

In double computing mode, overall dynamic logic switches $\overline{EN_0} \sim \overline{EN_3}$ are set to ground for pre-charging output to $VDD$ during each computing cycle.
As illustrated in Fig.~\ref{fig8:PIMMode} (b), both two paths could contribute current to the computational results.
For example, the left path in LPU opens to perform multiplication of $O_{ch0}[1]$ ($O_{ch0}[1] = w^c_{0,0}[1] \& INP)$ and the right path opens to perform multiplication of $O_{ch1}[1]$ ($O_{ch1}[1] = w^c_{0,1}[1] \& INN)$. 
With the support of the FCC algorithm and DBIS, two different inputs ($INN$ and $INP$) are broadcasted to LPU for two independent MAC operations.
In each compartment, only one row is activated every cycle to avoid read disturbance issues.
Bitwise \texttt{AND} operations are performed in all compartments simultaneously, and their results are accumulated vertically in the reconfigurable unit.

The reconfigurable unit consists of four adder units and a multiplexer (Mux) for flexibly combining the sum of the output elements.
Each adder unit comprises two adder trees, and each adder tree is responsible for accumulating \texttt{AND} results of $16$ compartments. 
The output of the adder unit is either directly extracted from two individual adder trees or the combination of the results from two adder trees.
The former output scheme represents two different output channels of $\cal O$ while the latter represents a single output channel. This flexibility allows for optimizing the performance and efficiency of the adder unit in different scenarios.

For \texttt{std-conv} and \texttt{pw-conv}, each adder unit accumulates the \texttt{AND} results in the same row of $32$ compartments by combining the results of two adder trees. For \texttt{dw-conv}, convolutional operations are performed on a per-channel basis without sharing $\cal I$ among filters.
Despite the employment of DBIS, the computational constraints dictate that only a pair of output channels can be concurrently computed.
To mitigate load latency, we load two pairs of twin-weights in one compartment row during each write operation. 
Consequently, we perform a two-stage computation and alternately activate two adder units during each stage to ensure the correctness of the results.
Moreover, for $3 \times 3 \times 1$ filters, the spatial utilization ratio of compartments is only $9/32$. To improve spatial utilization, each adder unit is responsible for computing two different channels and outputting the results simultaneously. 
The proposed methodology enables parallel computation of four output channels of $\cal O$ during \texttt{dw-conv} operation, which is analogous to the concurrent processing of \texttt{std-conv} and \texttt{pw-conv}.

\begin{figure}[t]
\centering
\includegraphics[width = 0.9\linewidth]{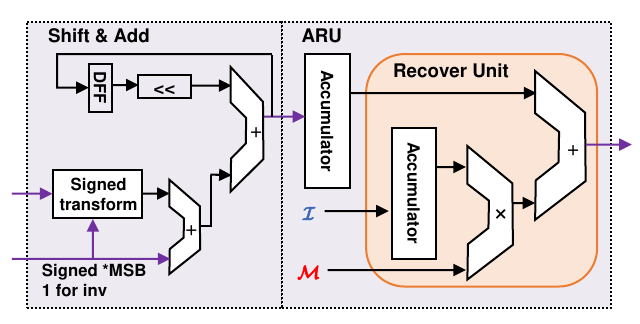}
\caption{Circuit design of merge unit.}
\label{fig9:ARU}
\end{figure}

\subsubsection{Merge unit}

Fig.~\ref{fig9:ARU} presents the circuit design of the merge unit that performs shift \& add operations and then recovers the convolution results. The partial results generated by the PIM core are first shifted and accumulated based on their respective bit position in shift \& add unit. 
Given that the \textit{Biased-Comp filters} are decomposed into $\cal M$ and \textit{Comp filters} during data mapping, as shown in Eq.~\ref{eq:eq7}, ARU is adopted to recover the final convolution results (${\cal O}$). 
In ARU, the Psums provided by the shift \& add unit are first accumulated in the vector-wise direction.
For \texttt{Conv} layers that are subjected to the FCC algorithm, the convolution results are obtained by assembling the accumulated Psums and the multiplication results ($\left(\sum \cal I \right) \times \cal M$). For \texttt{FC} layers, we disable the recover unit and only accumulate Psum for the final convolution results.
\begin{equation}\label{eq:eq7}
\begin{split}
    {\cal O}  &= \sum \left({\cal I}\ast f^{bc} \right) \\ 
              &=  \sum \left( {\cal I}\ast \left( f^{c} + {\cal M} \right)\right) \\ 
              &= \sum \left({\cal I}\ast f^{c} \right) +  \left( \sum {\cal I} \right)\times {\cal M} .
\end{split}
\end{equation}

\subsection{Data Mapping}\label{sec:Mapping}

\begin{figure}[t]
\centering
\includegraphics[width = 0.9\linewidth]{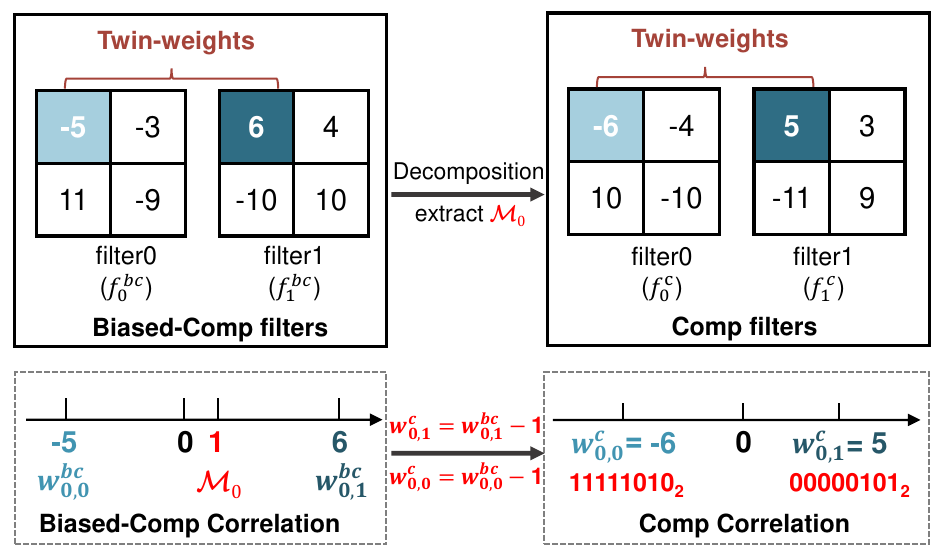}
\caption{Demonstration of decomposing Biased-Comp filters into Comp filters.}
\label{fig7:decomposition}
\end{figure}

To bridge the gap between FCC training algorithm and DDC-PIM architecture, we propose a versatile data mapping method, including offline data decomposition and mapping strategies for various kinds of computation.
Fig.~\ref{fig7:decomposition} presents how the \textit{Biased-Comp filters} obtained from FCC algorithm are decomposed into \textit{Comp filters} and $\cal M$. The cross-coupled structures are granted the capability of fully utilizing their complementary states.
For example, here $w^{bc}_{0,0} = -5$, $w^{bc}_{0,1} = 6$, and ${\cal M}_0 = 1$. After decomposition, $w^c_{0,0}$ and $w^c_{0,1}$ are $-6$ (i.e. $11111010_2$) and $5$ (i.e. $00000101_2$) respectively.  
The twin-weights in \textit{Comp filters} exhibit bitwise complementary properties, only one of them needs to be transferred.
Hence, we first extract half of the \textit{Comp filters} ($f^c_0, f^c_2, f^c_4, ...$), convert them into 1-dimensional vectors using the $im2col$ function, and splice every two \texttt{8 bit} vectors into a \texttt{16 bit} vector.

Due to the limited on-chip resources, these vectors and the corresponding mean values are transferred from off-chip DRAM to weight memory layer-by-layer. 
Each time a part of the weights is fetched from weight memory and loaded to PIM macros for performing matrix vector multiplication (MVM). 
Once these cached weights in weight memory are nearly exhausted for computation within the PIM macros, our system proactively pre-fetches the weights for the subsequent layer, effectively masking the latency typically associated with off-chip DRAM access.
For \texttt{std-conv} and \texttt{pw-conv}, DDC-PIM tiles these vectors in weight memory according to the total capacity of 4 PIM macros, and then splits them into sub-vectors based on the number of compartments.

For \texttt{dw-conv}, we adopt another mapping strategy with padding technique to solve the spatial under-utilization problem of the PIM core. 
For \texttt{FC} layer, we can regard it as a special case 
since we exclude it from the scope of the FCC algorithm.
We transfer all weights of the \texttt{FC} layer, tile
them according to the total capacity of 4 PIM macros, and compute in regular computing mode, as shown in Fig.~\ref{fig8:PIMMode}.



\subsubsection{Standard or pointwise convolution} 

\begin{figure}[t]
\centering
\includegraphics[width = 1.0\linewidth]{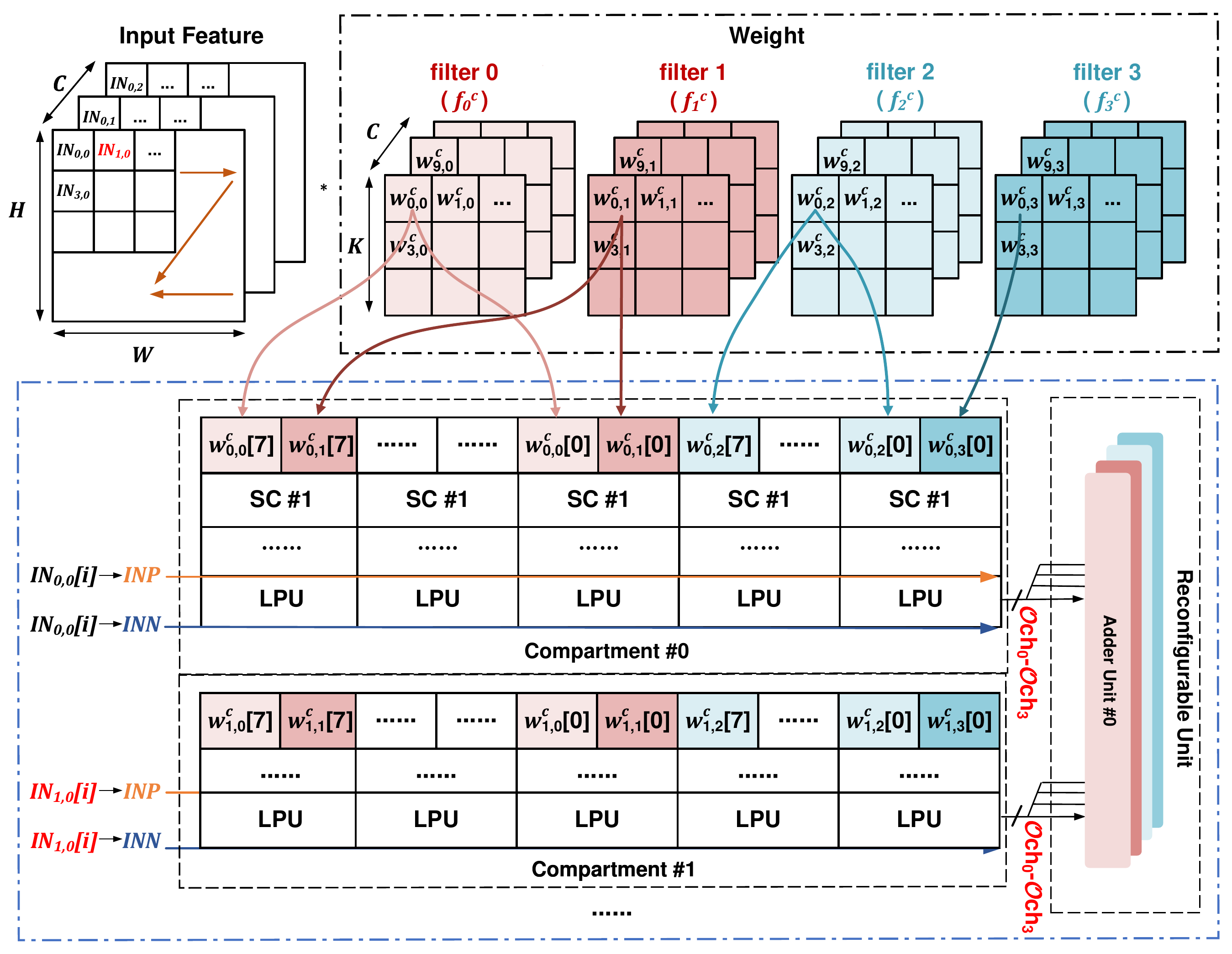}
\caption{Data mapping demonstration of \texttt{std-conv}.}
\label{fig10:std-mapping}
\end{figure}

For \texttt{std-conv} and \texttt{pw-conv}, two pairs of the twin-weights are mapped to the same row of the destined compartment.
Meanwhile, the weights within each sub-vector are allocated to rows with the same position across $32$ compartments.
Therefore, the results of \texttt{AND} operation between weights and vector-wise input are accumulated vertically.
We illustrate our method using a \texttt{std-conv} layer of 3D filters in Fig.~\ref{fig10:std-mapping}. 
By exploiting the bitwise complementary property of the twin-weights in $f^c_0$ and $f^c_1$, $f^c_2$ and $f^c_3$, we only transmit the corresponding weights in $f^c_0$ and $f^c_2$.
For the first row of Compartment \#$0$, we transform $w^c_{0,0}$ in $f^c_{0}$ and $w^c_{0,2}$ in $f^c_{2}$ into $\{w^c_{0,0},w^c_{0,2}\}$, and load them to this row in normal SRAM mode.
The data stored in first row from Compartment \#$0$ can then represent two pairs of twin-weights from four filters ($w^c_{0,0},w^c_{0,1},w^c_{0,2},w^c_{0,3}$), due to the cross-coupled structure of 6T SRAM and the bitwise complementary correlation ($w^c_{0,0}$$=$${\sim}$$w^c_{0,1}$ and $w^c_{0,2}$$=$${\sim}$$w^c_{0,3}$). 
In this case, the PIM core performs in double computing mode, where $INN$ and $INP$ receive the same vector-wise input, as different filters share the same $\cal I$ in \texttt{std-conv}.
Thus, the maximum computation parallelism ($X \times Y \times B$) supported by the DDC-PIM is $32 \times 4 \times 32$. Among them, $X$, $Y$, and $B$ denote the number of compartments, macros, and bits for parallel computing.

\subsubsection{Depthwise convolution}

\begin{figure}[t]
\centering
\includegraphics[width = 1.0\linewidth]{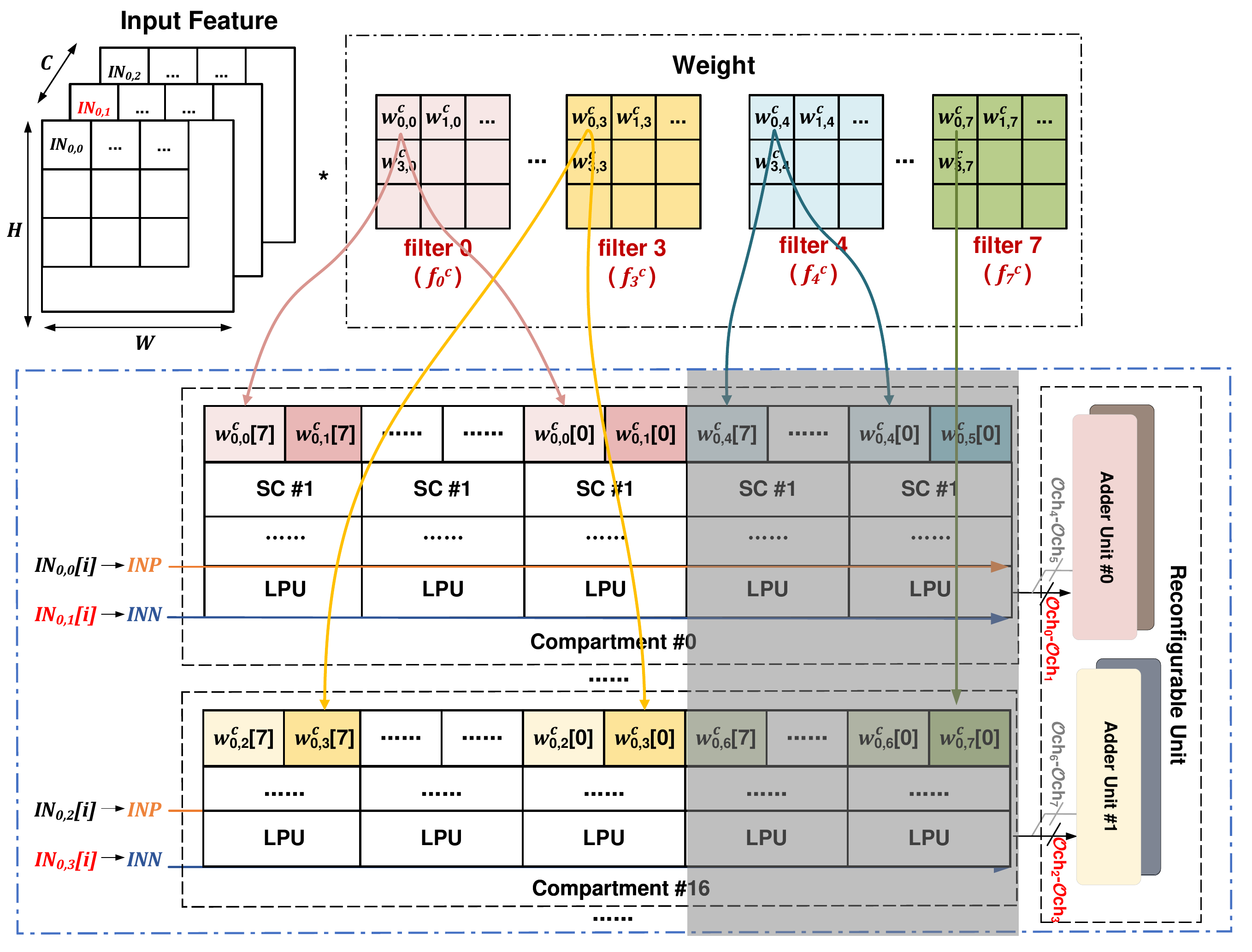}
\caption{Data mapping demonstration of \texttt{dw-conv}.}
\label{fig11:dw-mapping}
\end{figure}

Like \texttt{std-conv}, we apply filter transformation and assign two pairs of twin-weights to the same row of each compartment in \texttt{dw-conv}.
As we mentioned in the reconfigurable unit design in Section \ref{subsubsec:Architure}, only half of the compartments are activated during each computation stage.
We double the spatial utilization with the support of padding and reconfigurable unit.
For example, as shown in Fig.~\ref{fig11:dw-mapping}, we map $f^c_0 \sim f^c_3$ into Compartment \#$0$ $\sim$ Compartment \#$8$, and $f^c_4 \sim f^c_7$ into Compartment \#$16$ $\sim$ Compartment \#$24$.
When performing computation without the optimization techniques in DDC-PIM, the maximum computation parallelism is $9 \times 1 \times 8$ since only $9$ out of $32$ compartments are utilized. 
Based on the FCC algorithm and PIM core with DBIS, $INN$ and $INP$ can receive distinct vector-wise inputs, 
doubling the computation parallelism to $9 \times 1 \times 16$.
Meanwhile, the reconfigurable unit enables DDC-PIM to load four different filters ($f^c_{0},f^c_{2},f^c_{4},f^c_{6}$) simultaneously, representing eight filters ($f^c_{0}$ to $f^c_{7}$) for two alternating computations to further double the parallelism.
Thus, the overall maximum computation parallelism for \texttt{dw-conv} is $18 \times 1 \times 16$, equivalent to 4$\times$ acceleration.

\section{Evaluation Results}\label{sec:Experiments}

\subsection{Experimental Setup}\label{sec:setup}

\begin{figure}[t]
\centering
\includegraphics[width = 1.0\linewidth]{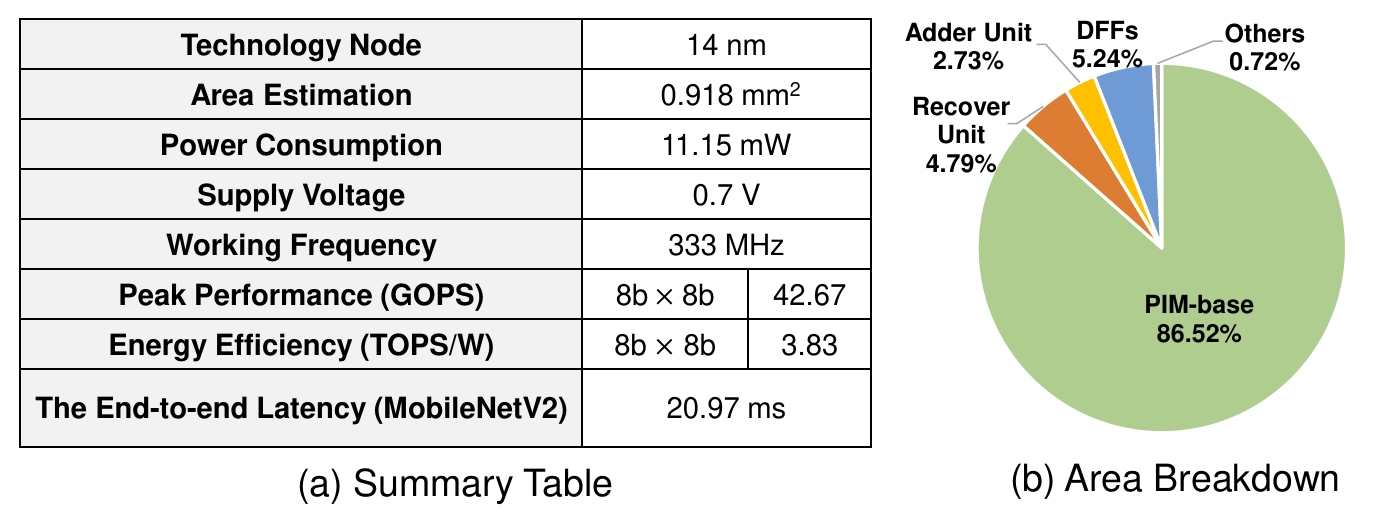}
\caption{Summary table of DDC-PIM and area breakdown of PIM macro.} 
\label{fig12:summary}
\end{figure} 

\textbf{Hardware implementation.} 
\hlblue{DDC-PIM is evaluated on $14$ nm technology, with a $128$ KB ping-pong memory, a $256$ KB weight memory, and four $4$ KB PIM macros.
The power consumption, latency, and area of PIM macros are extracted from the post-layout of customized design extension from \cite{yan20221}.
The area and power consumption of weight memory and ping-pong memory are estimated by PCACTI \cite{shafaei2014fincacti}.
The remaining digital modules are implemented with Verilog HDL and synthesized by Design Compiler, while the power consumption is obtained by PTPX.
Aiming to evaluate the performance of whole architecture and validate the data mapping, we implement a customized cycle-accurate C++ simulator and dataflow mapper.}

\begin{table*}[tp]
\caption{\hlblue{Comparison with prior works for PIM macros.}}
\label{tab:comparison}
\centering
\resizebox{\linewidth}{!}{
\begin{threeparttable}[b]
\begin{tabular}{|c|ccccc|ccc|}
\hline
PIM Macro Type & \multicolumn{5}{c|}{\textbf{Analog PIM}}  & \multicolumn{3}{c|}{\textbf{Digital PIM}}     \\ \hline
& \multicolumn{1}{c|}{\begin{tabular}[c]{@{}c@{}}Nat. Elec.'22 \cite{le202364}\\ \end{tabular}} & \multicolumn{1}{c|}{\begin{tabular}[c]{@{}c@{}}JETCAS'22 \cite{Garofalo2022AHI}\\ \end{tabular}} & \multicolumn{1}{c|}{\begin{tabular}[c]{@{}c@{}}Nat. Elec.’21 \cite{Hung2021AFC}\\ \end{tabular}} & \multicolumn{1}{c|}{\begin{tabular}[c]{@{}c@{}}VLSI'21 \cite{yin2021pimca}\\ (10T)\end{tabular}} & \begin{tabular}[c]{@{}c@{}}ISSCC'20 \cite{si202015}\\ (6T+LCC)\end{tabular} & \multicolumn{1}{c|}{\begin{tabular}[c]{@{}c@{}}ISSCC'21 \cite{chih202116}\\ (6T+LCC)\end{tabular}} &  \multicolumn{1}{c|}{\begin{tabular}[c]{@{}c@{}}ISSCC'22 \cite{yan20221}\\ (6T+LCC)\end{tabular}} & \textbf{This Work}    \\ \hline
PIM Device & \multicolumn{1}{c|}{PCM}& \multicolumn{1}{c|}{PCM} &\multicolumn{1}{c|}{RRAM} &\multicolumn{1}{c|}{SRAM}  & SRAM & \multicolumn{1}{c|}{SRAM}  & \multicolumn{1}{c|}{SRAM}  & SRAM  \\ \hline
Technology Node & \multicolumn{1}{c|}{14nm}& \multicolumn{1}{c|}{22nm}&\multicolumn{1}{c|}{22nm}& \multicolumn{1}{c|}{28nm}  & 28nm & \multicolumn{1}{c|}{22nm} &  \multicolumn{1}{c|}{28nm}  & 14nm  \\ \hline
\textbf{Array Size} & \multicolumn{1}{c|}{\textbf{64Kb}}& \multicolumn{1}{c|}{\textbf{64Kb}}&\multicolumn{1}{c|}{\textbf{4Mb}}& \multicolumn{1}{c|}{\textbf{3456Kb}} & \textbf{64Kb}& \multicolumn{1}{c|}{\textbf{64Kb}} &  \multicolumn{1}{c|}{\textbf{32Kb}}& \textbf{32Kb} \\ \hline
\textbf{Weight Capacity}& \multicolumn{1}{c|}{\textbf{64Kb}}&
\multicolumn{1}{c|}{\textbf{64Kb}}&\multicolumn{1}{c|}{\textbf{4Mb}}& \multicolumn{1}{c|}{\textbf{3456Kb}} & \textbf{64Kb} & \multicolumn{1}{c|}{\textbf{64Kb}} &  \multicolumn{1}{c|}{\textbf{32Kb}} & \textbf{64Kb} \\ \hline
Cell Type & \multicolumn{1}{c|}{8T4R}& \multicolumn{1}{c|}{/}&\multicolumn{1}{c|}{1T1R}& \multicolumn{1}{c|}{10T1C} & 6T & \multicolumn{1}{c|}{6T} &  \multicolumn{1}{c|}{6T} & 6T \\ \hline
Macro Area $(\text{mm}^2)$  & \multicolumn{1}{c|}{1.392}&
\multicolumn{1}{c|}{0.83}&\multicolumn{1}{c|}{6} & \multicolumn{1}{c|}{20.9} & 0.362  & \multicolumn{1}{c|}{0.202} &  \multicolumn{1}{c|}{0.040}  & 0.0115 \\ \hline
Integration Density$^{\ast}$ $(\text{Kb/mm}^2)$ & \multicolumn{1}{c|}{45.98@14nm}&
\multicolumn{1}{c|}{77.11@22nm}&\multicolumn{1}{c|}{682.67@22nm}& \multicolumn{1}{c|}{165.4@28nm} & 177@28nm & \multicolumn{1}{c|}{317@22nm}  &  \multicolumn{1}{c|}{800@28nm}  & 2783@14nm  \\ \hline
\begin{tabular}[c]{@{}c@{}}Integration Density $(\text{Kb/mm}^2)$\\ (Normalized to 28nm)\end{tabular}  & \multicolumn{1}{c|}{11.52}
&\multicolumn{1}{c|}{47.68}& \multicolumn{1}{c|}{422.09}   & \multicolumn{1}{c|}{165.4}  & 177  & \multicolumn{1}{c|}{196}         &  \multicolumn{1}{c|}{\textbf{800}}& \textbf{697} \\ \hline
\textbf{Weight Density$^{\dag}$ $(\text{Kb/mm}^2)$}   & \multicolumn{1}{c|}{\textbf{45.98@14nm}}   &
\multicolumn{1}{c|}{\textbf{77.11@22nm}}&\multicolumn{1}{c|}{\textbf{682.67@22nm}}  & \multicolumn{1}{c|}{\textbf{165.4@28nm}} & \textbf{177@28nm} & \multicolumn{1}{c|}{\textbf{317@22nm}}  &  \multicolumn{1}{c|}{\textbf{800@28nm}} & \textbf{5565@14nm}    \\ \hline
\textbf{\begin{tabular}[c]{@{}c@{}}Weight Density $(\text{Kb/mm}^2)$\\ (Normalized to 28nm)\end{tabular}}   & \multicolumn{1}{c|}{\textbf{11.52}}&
\multicolumn{1}{c|}{\textbf{47.68}}&\multicolumn{1}{c|}{\textbf{422.09 }} & \multicolumn{1}{c|}{\textbf{165.4}} & \textbf{177}  & \multicolumn{1}{c|}{\textbf{196}}  &  \multicolumn{1}{c|}{\textbf{800}} & \textbf{1391} \\ \hline
Computing Units & \multicolumn{1}{c|}{ADC}& \multicolumn{1}{c|}{ADC}& \multicolumn{1}{c|}{CW-CVS}& \multicolumn{1}{c|}{Flash ADC} & LMAR-SAR-ADC  & \multicolumn{1}{c|}{Logic Circuit} &  \multicolumn{1}{c|}{Logic Circuit} & Logic Circuit  \\ \hline
\textbf{\begin{tabular}[c]{@{}c@{}}Area Efficiency $(\text{GOPS/mm}^2)$\\ (Normalized to 28nm)\end{tabular}} & \multicolumn{1}{c|}{\textbf{177.38/63 (8b/8b)}}&
\multicolumn{1}{c|}{\textbf{ 712.15 (8b/4b)}}&\multicolumn{1}{c|}{\textbf{3.47 (8b/8b)}}& \multicolumn{1}{c|}{\textbf{234 (1b/1b)}} & \textbf{84.2 (8b/8b)} & \multicolumn{1}{c|}{\textbf{2802.5 (8b/8b)}}  &  \multicolumn{1}{c|}{\textbf{133.3 (8b/8b)}}  & \textbf{231.9 (8b/8b)} \\ \hline
\textbf{Energy Efficiency $(\text{TOPS/W})$} & \multicolumn{1}{c|}{\textbf{9.76/2.48 (8b/8b)}}&\multicolumn{1}{c|}{\textbf{6.39 (8b/4b)}}& \multicolumn{1}{c|}{\textbf{15.60 (8b/8b)}}& \multicolumn{1}{c|}{\textbf{588 (1b/1b)}} & \textbf{14.1 (8b/8b)} & \multicolumn{1}{c|}{\textbf{24.7 (8b/8b)}} &  \multicolumn{1}{c|}{\textbf{27.38 (8b/8b)}} & \textbf{72.41 (8b/8b)} \\ \hline
\end{tabular}
\begin{tablenotes}
    \item $^{\ast}$ Integration Density = Array Size / Macro Area 
    \item $^{\dag}$ Weight Density = Weight Capacity / Macro Area
\end{tablenotes}
\end{threeparttable}
}
\end{table*}

\textbf{PIM baseline.} 
In order to assess the performance improvement achieved by our algorithm/architecture co-design, we implement a digital PIM as a baseline.
Compared with our DDC-PIM, the baseline design does not include reconfigurable unit, dual-broadcast input structure and recover unit. 
Furthermore, the PIM core in the baseline design is constrained to only operate in regular computing mode.
The rest of hardware settings are identical to those of our DDC-PIM architecture.

\textbf{Benchmarks and models.} 
 We evaluate two popular compact NN models, MobileNetV2 and EfficientNet-B0, on the CIFAR10 dataset benchmark.  
In addition, to demonstrate the generality and applicability of the FCC algorithm across different neural network architectures, we also perform experiments on three representative regular NN models, namely AlexNet, VGG19 and ResNet18. 
All of these models are trained for 1000 epochs. We apply \texttt{INT8} quantization on inputs and weights for all layers.

\subsection{DDC-PIM Implementation Summary}\label{sec:summary}
\hlblue{As shown in Fig.~\ref{fig12:summary}(a), the total area and power of DDC-PIM are about $0.918$ $\text{mm}^2$ and $11.15$ mW.
The clock frequency is $333$ MHz and the peak performance is about $42.67$ GOPS at $\text{8b} \times \text{8b}$. The end-to-end latency of MobileNetV2 is $20.97$ ms, with the MVM operations exhibiting a latency of $18.02$ ms.
The PIM macro area breakdown analysis is shown in Fig.~\ref{fig12:summary}(b).
PIM macro can be divided into base digital PIM logic (PIM-base), additional logic introduced by our techniques (DFFs, adder units and the recover unit), and others. PIM-base in PIM macro is equivalent to the counterpart in \cite{yan20221}, and the additional logic is designed for supporting our co-design.
In this work, the complementary states ($Q/\overline{Q}$) in 6T SRAM represent two individual bits for parallel computation. Hence, compared with PIM-base in \cite{yan20221}, DDC-PIM needs extra storage units (DFFs) and computation units (adder units) to process the additional information. These units only incur a minor area overhead of about $5.24\%$ and $2.73\%$, respectively.
Meanwhile, DDC-PIM decomposes \textit{Biased-Comp filters} into \textit{Comp filters} and $\cal M$, which requires extra computation units to recover the final results, costing only about $4.79\%$ overhead in area consumption. The energy efficiency in our PIM macro is $72.41$ TOPS/W at $8$b $\times$ $8$b. }

\subsection{Comparison with PIM Macros in Prior Works}

\hlblue{Tab. \ref{tab:comparison} presents a comprehensive comparison of PIM macros among DDC-PIM and other state-of-the-art studies, which can be categorized into the analog domain \cite{yin2021pimca, si202015,Garofalo2022AHI,le202364,Hung2021AFC}  and digital domain \cite{yan20221,chih202116}.}
We mainly focus on comparing weight capacity, integration density, weight density, and area efficiency of DDC-PIM macro against the others.
Generally, the weight capacity of a PIM macro is equal to its array size, thus its integration density and weight density are also the same.
For example, the array size and weight capacity of \cite{yin2021pimca} are both $3456$ Kb.
The integration density and weight density are about $165.4$ Kb/mm$^2$ at $28$ nm.  
Meanwhile, with a fixed process and array size, the weight density of analog PIM is usually much lower than digital PIM, due to the extra area overhead introduced by ADC/DAC.
For example, the integration density of \cite{yan20221} and \cite{chih202116} are much larger than that of \cite{yin2021pimca} and \cite{si202015}. 
In this work, by exploiting the complementary state pairs as independent bits of information, the weight capacity of DDC-PIM is twice its array size.
To facilitate a fair comparison, we scale both the integration density and weight density to $28$ nm technology node.
As PIM-base in PIM macro is equivalent to the counterpart in \cite{yan20221}, the additional logic for supporting our co-design brings a slight decrease in integration density.
Meanwhile, due to that a pair of computing units in DDC-PIM can support two independent \texttt{AND} operations, the area efficiency is also improved by about $1.74\times$ compared with \cite{yan20221}. 

\begin{figure}[t]
\centering
\includegraphics[width = 0.9\linewidth]{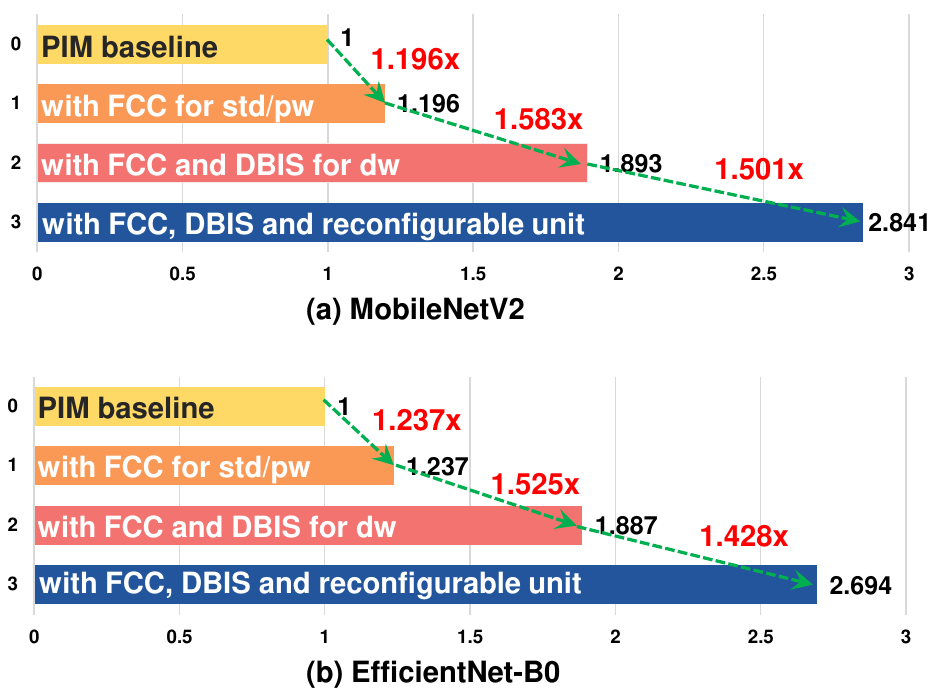}
\caption{Speedup analysis for MobileNetV2 and EfficientNet-B0. Here we illustrate the speedup of the FCC algorithm for \texttt{std-conv} and \texttt{pw-conv}, FCC algorithm for \texttt{dw-conv} with DBIS and reconfigurable unit over PIM baseline.}
\label{fig13:speedup_analysis}
\end{figure} 

\begin{table*}[t]
\caption{Accuracy evaluation of FCC algorithm applied on different layers and different models on CIFAR10 dataset.}
\label{tab:comparison for regular NN}
\centering
\begin{tabular}{|cc|cccccc|}
\hline
\multicolumn{2}{|c|}{\multirow{3}{*}{Model}}  & \multicolumn{1}{c|}{FCC Not Applied} & \multicolumn{2}{c|}{FCC Applied on \texttt{Conv} Layers} & \multicolumn{2}{c|}{FCC Applied on both \texttt{Conv} and \texttt{FC}} & \multicolumn{1}{c|}{\multirow{3}{*}{\begin{tabular}[c]{@{}c@{}}Param. Ratio \\ of \texttt{FC} Layers\end{tabular}}} \\ \cline{3-7} 
\multicolumn{2}{|c|}{}  & \multicolumn{1}{c|}{\begin{tabular}[c]{@{}c@{}}Top-1 Accu.\\ Baseline (\%)\end{tabular}} & \multicolumn{1}{c|}{\begin{tabular}[c]{@{}c@{}}Top-1 Accu. for \\ \texttt{Conv} Layers (\%)\end{tabular}} & \multicolumn{1}{c|}{\begin{tabular}[c]{@{}c@{}}Accu. Drop for \\ \texttt{Conv} Layers (\%)\end{tabular}} & \multicolumn{1}{c|}{\begin{tabular}[c]{@{}c@{}}Top-1 Accu. for \\  \texttt{Conv} \& \texttt{FC} (\%)\end{tabular}} & \multicolumn{1}{c|}{\begin{tabular}[c]{@{}c@{}}Accu. Drop for \\ \texttt{Conv} \& \texttt{FC} (\%)\end{tabular}} &  \\ \hline
\multicolumn{1}{|c|}{\multirow{2}{*}{\begin{tabular}[c]{@{}c@{}}Compact \\ NNs\end{tabular}}} & MobileNetV2  & \multicolumn{1}{c|}{96.71}  & \multicolumn{1}{c|}{95.99}  & \multicolumn{1}{c|}{0.72}  & \multicolumn{1}{c|}{95.69}  & \multicolumn{1}{c|}{1.02}  & 0.57\%  \\ \cline{2-8} 
\multicolumn{1}{|c|}{}  & Efficient-B0 & \multicolumn{1}{c|}{92.77}  & \multicolumn{1}{c|}{91.65}  & \multicolumn{1}{c|}{1.12}  & \multicolumn{1}{c|}{90.87}  & \multicolumn{1}{c|}{1.90}  & 0.11\%  \\ \hline
\multicolumn{1}{|c|}{\multirow{3}{*}{\begin{tabular}[c]{@{}c@{}}Regular\\ NNs\end{tabular}}}  & AlexNet      & \multicolumn{1}{c|}{93.08} & \multicolumn{1}{c|}{92.52}  & \multicolumn{1}{c|}{0.56}  & \multicolumn{1}{c|}{91.20}  & \multicolumn{1}{c|}{1.88} & 79.12\%  \\ \cline{2-8}
\multicolumn{1}{|c|}{} & VGG19  & \multicolumn{1}{c|}{96.29} & \multicolumn{1}{c|}{95.64}  & \multicolumn{1}{c|}{0.65}  & \multicolumn{1}{c|}{95.11}  & \multicolumn{1}{c|}{1.18}  & 55.71\%   \\ \cline{2-8}  
\multicolumn{1}{|c|}{}   & ResNet18     & \multicolumn{1}{c|}{97.15} & \multicolumn{1}{c|}{96.73}  & \multicolumn{1}{c|}{0.42} & \multicolumn{1}{c|}{95.97}  & \multicolumn{1}{c|}{1.18}  & 0.04\%  \\ \hline
\end{tabular}
\end{table*}

\subsection{Speedup for MobileNetV2 and EfficientNet-B0}

Fig.~\ref{fig13:speedup_analysis} clearly illustrates the acceleration in MobileNetV2 and EfficientNet-B0 gained by co-designing the FCC algorithm, data mapping, and architecture.
By applying the FCC algorithm to both \texttt{std-conv} and \texttt{pw-conv}, we achieve a speedup of about $1.196\times$ for MobileNetV2 and $1.237\times$ for EfficientNet-B0, respectively. Moreover, the combination of the FCC algorithm and DBIS improves the efficiency of \texttt{dw-conv}, achieving a speedup of $1.583\times$ for MobileNetV2 and $1.525\times$ for EfficientNet-B0. Furthermore, the proposed DDC-PIM architecture, which incorporates the FCC algorithm, DBIS, and a reconfigurable unit, attains a speedup of about $1.501\times$ for MobileNetV2 and $1.428\times$ for EfficientNet-B0.
For \texttt{dw-conv}, the speedup brought by the FCC algorithm solely relies on the support of DBIS, thus these two approaches must be bonded during execution.
Although \texttt{dw-conv} has fewer parameters and fewer computation requirements than \texttt{pw-conv} and \texttt{std-conv}, the overall latency of compact NNs is still dominated by \texttt{dw-conv} due to its low computation parallelism.
With the optimization techniques tailored for \texttt{dw-conv}, DDC-PIM can achieve about $2.841\times$ and $2.694\times$ speedup for MobileNetV2 and EfficientNet-B0 respectively.


\subsection{Evaluation of FCC Algorithm}

\begin{figure}[t]
\centering
\includegraphics[width = 0.9\linewidth]{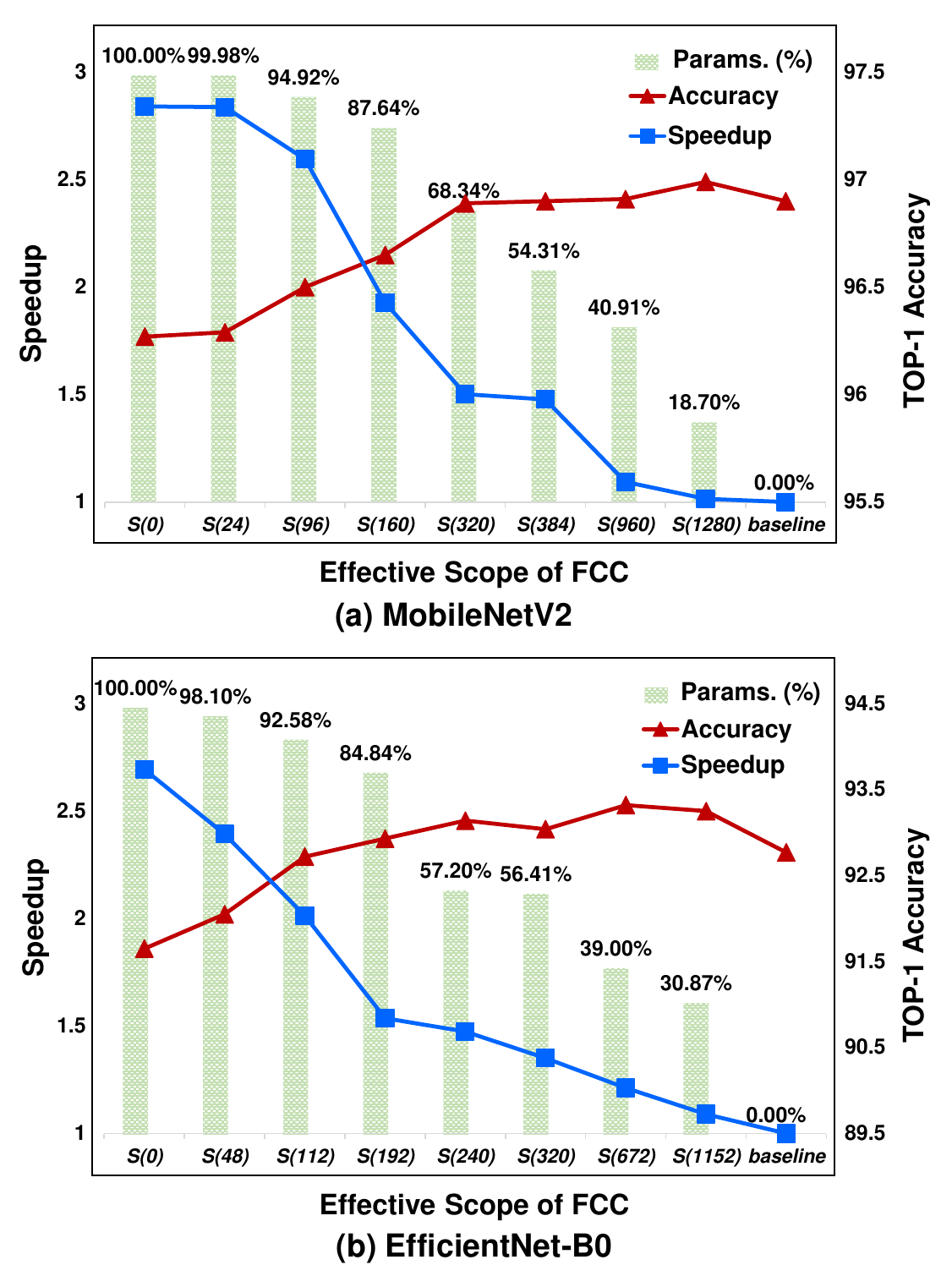}
\caption{Speedup and accuracy tradeoff for FCC algorithm on MobileNetV2 and EfficientNet-B0. The height of bars in the bar diagram denotes the proportion of parameters within $S(i)$ to the overall parameters.}
\label{fig14:diff_layers}
\end{figure} 

Through a comprehensive set of experiments, we demonstrate that the FCC algorithm has a heterogeneous impact on the performance of NN model’s layers, indicating a varying degree of susceptibility across different levels of abstraction. This impact translates into distinct acceleration effects. For simplicity, we introduce the effective scope $S(i)$ as a set composed of all the layers in a given model with
more than $i$ filters, and apply the FCC algorithm to the layers in $S(i)$. Therefore, we assess the tradeoff between speedup and accuracy by altering the $S(i)$ of the FCC algorithm and present the results in Fig.~\ref{fig14:diff_layers}. In this case, all layers represent all \texttt{Conv} layers. 
The results show that when FCC is applied to all layers, DDC-PIM can obtain $2.841\times$ and $2.694\times$ speedup with only $0.72\%$ and $1.12\%$ accuracy drop for MobileNetV2 and EfficientNet-B0 respectively.
To gain more insights into the benefits of our approach, we scrutinize the number of parameters, accuracy, and speedup at $S(112)$ in Fig.~\ref{fig14:diff_layers}(b). DDC-PIM can apply FCC to of most $92.58\%$ parameters, achieving a $2.01\times$ speedup without any accuracy degradation.
Furthermore, we observe that applying the FCC algorithm to a smaller $S(i)$ may result in a deterioration of the inference accuracy.
However, applying the FCC algorithm to a larger $S(i)$ may exceed the baseline accuracy. For applications with stringent requirements on model accuracy, we can investigate a layered application of the FCC algorithm. 

To demonstrate the generality of the FCC algorithm, we also perform experiments on regular NNs such as AlexNet, VGG19, and ResNet18. Tab. \ref{tab:comparison for regular NN} illustrates the versatility and robustness of FCC algorithm applied on different layers and NN models.
First, we compare the accuracy of applying the FCC algorithm only to \texttt{Conv} layers as well as applying it to both the \texttt{Conv} layers and \texttt{FC} layers.
The findings indicate that \texttt{FC} layers are more susceptible to the FCC algorithm than the \texttt{Conv} layers, as the former setting yields higher accuracy than the latter.
Moreover, the results imply that the FCC algorithm can handle different levels of complexity in the NNs. Generally, regular NN models exhibit a higher degree of compatibility with the FCC algorithm with lower accuracy degradation due to its higher redundancy.
However, due to \texttt{FC} layers accounting for a significant proportion of the total number of parameters in regular NNs, applying the FCC algorithm both to \texttt{Conv} layers and \texttt{FC} layers obtains larger accuracy deterioration than in compact NNs. Taking
AlexNet as an example, the accuracy drop for \texttt{Conv} layers is $0.56\%$, but for \texttt{Conv} layers and \texttt{FC} layers is up to $1.88\%$.
Although applying the FCC algorithm to all layers, including the \texttt{Conv} layers and \texttt{FC} layers, can achieve better acceleration. However, this often comes at the expense of significant accuracy degradation. Hence, the choice of FCC algorithms should be based on the trade-off between acceleration and accuracy requirements.

\section{Discussions}\label{sec:Discussion}

\textbf{Existing pruning methods.} 
\hlblue{Apart from designing compact NNs, network sparsification is another promising compression algorithm to reduce data and computation requirements. 
Numerous effective methods have been proposed to introduce sparsity \cite{deng2020model, yang2021s, dai2020sparsetrain}. 
Weight pruning, indeed, holds a prominent position among these methods.
Generally, weight pruning can be divided into fine-grained pruning and coarse-grained pruning. 
Fine-grained pruning allows for precise removal of unnecessary weights, leading to higher compression rates compared to coarse-grained pruning. 
However, pruning individual weights may introduce irregularities in the network, leading to increased computational overhead during inference due to non-contiguous memory access or additional indexing operations.
Coarse-grained pruning achieves significant computational savings by removing entire neurons, channels, or layers from a neural network. 
However, it will introduce non-negligible accuracy loss.
To further reduce memory access and computation cost, some works apply the joint-way compression with multiple approaches. 
A usual case is using compact models such as MobileNet as the base model to do additional compression such as quantization or sparsification. 
However, compact models are already size-efficient than normal models, which usually impedes further model compression. 
For example, the weight pruning on MobileNet can achieve only $50\% \sim 60\%$ sparsity without paying significant accuracy loss \cite{deng2020model}.}

\begin{table}[tp]
\caption{\hlblue{Comparison of accuracy and compression ratio of MobileNetV2 with different methods on CIFAR-100.}}
\label{tab:Orthogonal}
\centering
\begin{tabular}{|c|c|c|c|c|}
\hline
Model    & Method    & \begin{tabular}[c]{@{}c@{}}Top-1\\ Accu.\\ (\%)\end{tabular} & \begin{tabular}[c]{@{}c@{}}Accu.\\ Drop\\ (\%)\end{tabular} & \begin{tabular}[c]{@{}c@{}}Compression\\ Ratio \end{tabular} \\ \hline
\multirow{4}{*}{MobileNetV2} & Original     & $80.48$  & $0$   & $0\%$    \\ \cline{2-5} 
                             & $2\colon4$ Pruning     & $79.94$   & $0.54$  & $50\%$  \\ \cline{2-5} 
                             & \begin{tabular}[c]{@{}c@{}}FCC algorithm \\ with $2\colon4$ Pruning\end{tabular} & $78.81$ & $1.13$ & $\sim75\%$  \\ \hline
\end{tabular}
\end{table}

\textbf{Apply FCC algorithm on pruned models.} 
\hlblue{FCC algorithm is a novel compression method that is designed to reduce the memory access and mitigate the effects of computational irregularity. 
The compatibility between the FCC algorithm and traditional pruning techniques is an intriguing question that merits further investigation. 
In order to thoroughly examine and confirm this compatibility, we have conducted additional supplementary experiments. 
We choose the well-known fine-grained $2\colon4$ structured pruning algorithm proposed by NVIDIA to verify the orthogonal nature of our algorithm relative to traditional pruning techniques \cite{mishra2021accelerating}.
Specifically, we have employed the $2\colon4$ pruning algorithm proposed by NVIDIA with an accuracy of $79.94\%$ and achieved a $50\%$ compression ratio, as shown in Tab.~\ref{tab:Orthogonal}. 
Then, we have further conducted experiments incorporating the FCC algorithm with the $2\colon4$ pruning algorithm. 
The experimental accuracy is $78.81\%$. 
This result demonstrates the compatibility between the FCC algorithm and the $2\colon4$ structured pruning technique. 
This compression method achieves an additional compression of nearly $50\%$ on top of the initial $50\%$ achieved by the original $2\colon4$ pruning algorithm, without introducing significant accuracy loss.}

\textbf{Applicability for other DNNs.} 
\hlblue{
To verify the applicability of the FCC algorithm, we choose MobileViT-XS, a variant of lightweight transformer models, for supplementary experiments. 
The experimental results, as shown in Tab.~\ref{tab:OTHER-NN}, illustrate that applying
the FCC algorithm to the convolutional layer of MobileViT-XS will not introduce a large accuracy loss.
Exploring the potential application of our algorithm to alternative neural networks constitutes a prospective avenue of our research. 
We plan to conduct a more comprehensive analysis in the future to provide additional perspectives on the advantages and constraints of our approach within a more expansive framework.}

\begin{table}[tp]
\caption{\hlblue{Accuracy comparison of MobileViT-XS on CIFAR-10.}}
\label{tab:OTHER-NN}
\begin{tabular}{|c|c|c|}
\hline
Model                        & Method        & Top-1 Accu.(\%) \\ \hline
\multirow{2}{*}{MobileViT-XS} & Original   & $90.88$     \\ \cline{2-3} 
                              & \begin{tabular}[c]{@{}c@{}}FCC algorithm for \texttt{Conv} layers\end{tabular} & $89.04$                                         \\ \hline
\end{tabular}
\end{table}

\section{Conclusions}\label{sec:Conclusion}
This paper presents DDC-PIM, a novel co-design of algorithm and architecture that enhances the data capacity of SRAM without altering its structure.
At the algorithm level, the proposed FCC algorithm leverages the complementary property of filters to transform adjacent filter pairs into the bitwise complementary format with negligible accuracy loss.
The DDC-PIM architecture exploits the intrinsic cross-coupled structures of 6T SRAM cells to double the equivalent data capacity. 
Moreover, DDC-PIM adopts a flexible data mapping method that adapts to various convolution operations. \hlblue{Experimental results demonstrate that DDC-PIM 
yields about $2.84\times$ and $2.69\times$ speedup on MobileNetV2 and EfficientNet-B0, compared with  baseline implementations, and can increase weight density and area efficiency up to $8.41\times$ and $2.75\times$, compared with state-of-the-art SRAM-based PIM.}

{
\small
\bibliographystyle{IEEEtran}

\begin{thebibliography}{10}
\providecommand{\url}[1]{#1}
\csname url@samestyle\endcsname
\providecommand{\newblock}{\relax}
\providecommand{\bibinfo}[2]{#2}
\providecommand{\BIBentrySTDinterwordspacing}{\spaceskip=0pt\relax}
\providecommand{\BIBentryALTinterwordstretchfactor}{4}
\providecommand{\BIBentryALTinterwordspacing}{\spaceskip=\fontdimen2\font plus
\BIBentryALTinterwordstretchfactor\fontdimen3\font minus
  \fontdimen4\font\relax}
\providecommand{\BIBforeignlanguage}[2]{{%
\expandafter\ifx\csname l@#1\endcsname\relax
\typeout{** WARNING: IEEEtran.bst: No hyphenation pattern has been}%
\typeout{** loaded for the language `#1'. Using the pattern for}%
\typeout{** the default language instead.}%
\else
\language=\csname l@#1\endcsname
\fi
#2}}
\providecommand{\BIBdecl}{\relax}
\BIBdecl

\bibitem{krizhevsky2017imagenet}
A.~Krizhevsky, I.~Sutskever, and G.~E. Hinton, ``{ImageNet Classification with
  Deep Convolutional Neural Networks},'' \emph{Communications of the ACM},
  vol.~60, no.~6, pp. 84--90, 2017.

\bibitem{szegedy2017inception}
C.~Szegedy, S.~Ioffe, V.~Vanhoucke, and A.~Alemi, ``{Inception-v4,
  inception-resnet and the impact of residual connections on learning},'' in
  \emph{Proceedings of the Association for the Advancement of Artificial
  Intelligence (AAAI)}, 2017.

\bibitem{he2016deep}
K.~He, X.~Zhang, S.~Ren, and J.~Sun, ``{Deep residual learning for image
  recognition},'' in \emph{Proceedings of the Conference on Computer Vision and
  Pattern Recognition (CVPR)}, 2016.

\bibitem{burchi2023audio}
M.~Burchi and R.~Timofte, ``{Audio-Visual Efficient Conformer for Robust Speech
  Recognition},'' in \emph{Proceedings of the Winter Conference on Applications
  of Computer Vision (WACV)}, 2023.

\bibitem{zhang2017very}
Y.~Zhang, W.~Chan, and N.~Jaitly, ``{Very deep convolutional networks for
  end-to-end speech recognition},'' in \emph{Proceedings of the International
  Conference on Acoustics, Speech and Signal Processing (ICASSP)}, 2017.

\bibitem{wang2022internimage}
W.~Wang, J.~Dai, Z.~Chen, Z.~Huang, Z.~Li, X.~Zhu, X.~Hu, T.~Lu, L.~Lu, H.~Li
  \emph{et~al.}, ``Internimage: Exploring large-scale vision foundation models
  with deformable convolutions,'' in \emph{Proceedings of the IEEE/CVF
  Conference on Computer Vision and Pattern Recognition (CVPR)}, 2023, pp.
  14\,408--14\,419.

\bibitem{fang2022eva}
Y.~Fang, W.~Wang, B.~Xie, Q.~Sun, L.~Wu, X.~Wang, T.~Huang, X.~Wang, and
  Y.~Cao, ``Eva: Exploring the limits of masked visual representation learning
  at scale,'' in \emph{Proceedings of the IEEE/CVF Conference on Computer
  Vision and Pattern Recognition (CVPR)}, 2023, pp. 19\,358--19\,369.

\bibitem{su2022towards}
W.~Su, X.~Zhu, C.~Tao, L.~Lu, B.~Li, G.~Huang, Y.~Qiao, X.~Wang, J.~Zhou, and
  J.~Dai, ``Towards all-in-one pre-training via maximizing multi-modal mutual
  information,'' in \emph{Proceedings of the IEEE/CVF Conference on Computer
  Vision and Pattern Recognition (CVPR)}, 2023, pp. 15\,888--15\,899.

\bibitem{howard2017mobilenets}
A.~G. Howard, M.~Zhu, B.~Chen, D.~Kalenichenko, W.~Wang, T.~Weyand,
  M.~Andreetto, and H.~Adam, ``{MobileNets: Efficient Convolutional Neural
  Networks for Mobile Vision Applications},'' \emph{arXiv preprint
  arXiv:1704.04861}, 2017.

\bibitem{tan2019efficientnet}
M.~Tan and Q.~Le, ``{EfficientNet: Rethinking Model Scaling for Convolutional
  Neural Networks},'' in \emph{Proceedings of the International Conference on
  Machine Learning (ICML)}, 2019.

\bibitem{yin2021pimca}
S.~Yin, B.~Zhang, M.~Kim, J.~Saikia, S.~Kwon, S.~Myung, H.~Kim, S.~J. Kim,
  M.~Seok, and J.-s. Seo, ``{PIMCA: A 3.4-Mb Programmable In-Memory Computing
  Accelerator in 28nm for On-Chip DNN Inference},'' in \emph{Proceedings of the
  Symposium on VLSI Circuits (VLSIC)}, 2021.

\bibitem{gonugondla201842pj}
S.~K. Gonugondla, M.~Kang, and N.~Shanbhag, ``{A 42pJ/decision 3.12TOPS/W
  Robust In-Memory Machine Learning Classifier with On-Chip Training},'' in
  \emph{Proceedings of the International Solid-State Circuits Conference
  (ISSCC)}, 2018.

\bibitem{su202015}
J.-W. Su, X.~Si, Y.-C. Chou, T.-W. Chang, W.-H. Huang, Y.-N. Tu, R.~Liu, P.-J.
  Lu, T.-W. Liu, J.-H. Wang \emph{et~al.}, ``{A 28nm 64Kb Inference-Training
  Two-Way Transpose Multibit 6T SRAM Compute-In-Memory Macro for AI Edge
  Chips},'' in \emph{Proceedings of the International Solid-State Circuits
  Conference (ISSCC)}, 2020.

\bibitem{yan20221}
B.~Yan, J.-L. Hsu, P.-C. Yu, C.-C. Lee, Y.~Zhang, W.~Yue, G.~Mei, Y.~Yang,
  Y.~Yang, H.~Li \emph{et~al.}, ``{A 1.041-Mb/mm$^2$ 27.38-TOPS/W Signed-INT8
  Dynamic-Logic-Based ADC-less SRAM Compute-In-Memory Macro in 28nm with
  Reconfigurable Bitwise Operation for AI and Embedded Applications},'' in
  \emph{Proceedings of the International Solid-State Circuits Conference
  (ISSCC)}, 2022.

\bibitem{Imani2019DigitalPIMDP}
M.~Imani, S.~Gupta, Y.~Kim, M.~Zhou, and T.~Rosing, ``{DigitalPIM:
  Digital-based Processing In-Memory for Big Data Acceleration},'' in
  \emph{Proceedings of Great Lakes Symposium on VLSI (GLSVLSI)}, 2019.

\bibitem{8980299}
M.~Imani, S.~Gupta, Y.~Kim, and T.~Rosing, ``{FloatPIM: In-Memory Acceleration
  of Deep Neural Network Training with High Precision},'' in \emph{Proceedings
  of International Symposium on Computer Architecture (ISCA)}, 2019.

\bibitem{chen2022accelerating}
X.~Chen, X.~Wang, X.~Jia, J.~Yang, G.~Qu, and W.~Zhao, ``{Accelerating
  Graph-Connected Component Computation with Emerging Processing-In-Memory
  Architecture},'' \emph{Transactions on Computer-Aided Design of Integrated
  Circuits and Systems (TCAD)}, vol.~41, no.~12, pp. 5333--5342, 2022.

\bibitem{Zhang2021TimeDomainCI}
Y.~Zhang, J.~Wang, C.~Lian, Y.~Bai, G.~Wang, Z.~Zhang, Z.~Zheng, L.~Chen,
  K.~Zhang, G.~Sirakoulis \emph{et~al.}, ``{Time-Domain Computing in Memory
  Using Spintronics for Energy-Efficient Convolutional Neural Network},''
  \emph{Transactions on Circuits and Systems I: Regular Papers (TCAS-I)},
  vol.~68, no.~3, pp. 1193--1205, 2021.

\bibitem{9218660}
X.~Wang, J.~Yang, Y.~Zhao, Y.~Qi, M.~Liu, X.~Cheng, X.~Jia, X.~Chen, G.~Qu, and
  W.~Zhao, ``{TCIM: Triangle Counting Acceleration With Processing-In-MRAM
  Architecture},'' in \emph{Proceedings of the Design Automation Conference
  (DAC)}, 2020.

\bibitem{9729451}
T.~Kim, Y.~Jang, M.-G. Kang, B.-G. Park, K.-J. Lee, and J.~Park, ``{SOT-MRAM
  Digital PIM Architecture With Extended Parallelism in Matrix
  Multiplication},'' \emph{Transactions on Computers (TC)}, vol.~71, no.~11,
  pp. 2816--2828, 2022.

\bibitem{kang2018multi}
M.~Kang, S.~K. Gonugondla, A.~Patil, and N.~R. Shanbhag, ``{A Multi-Functional
  In-Memory Inference Processor Using a Standard 6T SRAM Array},''
  \emph{Journal of Solid-State Circuits (JSSC)}, vol.~53, no.~2, pp. 642--655,
  2018.

\bibitem{sinangil20207}
M.~E. Sinangil, B.~Erbagci, R.~Naous, K.~Akarvardar, D.~Sun, W.-S. Khwa, H.-J.
  Liao, Y.~Wang, and J.~Chang, ``{A 7-nm Compute-In-Memory SRAM Macro
  Supporting Multi-Bit Input, Weight and Output and Achieving 351 TOPS/W and
  372.4 GOPS},'' \emph{Journal of Solid-State Circuits (JSSC)}, vol.~56, no.~1,
  pp. 188--198, 2020.

\bibitem{zhang2018recryptor}
Y.~Zhang, L.~Xu, Q.~Dong, J.~Wang, D.~Blaauw, and D.~Sylvester, ``{Recryptor: A
  Reconfigurable Cryptographic Cortex-M0 Processor With In-Memory and
  Near-Memory Computing for IoT Security},'' \emph{Journal of Solid-State
  Circuits (JSSC)}, vol.~53, no.~4, pp. 995--1005, 2018.

\bibitem{si202015}
X.~Si, Y.-N. Tu, W.-H. Huang, J.-W. Su, P.-J. Lu, J.-H. Wang, T.-W. Liu, S.-Y.
  Wu, R.~Liu, Y.-C. Chou \emph{et~al.}, ``{A 28nm 64Kb 6T SRAM
  Computing-in-Memory Macro with 8b MAC Operation for AI Edge Chips},'' in
  \emph{Proceedings of the International Solid-State Circuits Conference
  (ISSCC)}, 2020.

\bibitem{yue2022br}
Z.~Yue, Y.~Wang, Y.~Qin, L.~Liu, S.~Wei, and S.~Yin, ``{BR-CIM: An Efficient
  Binary Representation Computation-In-Memory Design},'' \emph{Transactions on
  Circuits and Systems I: Regular Papers (TCAS-I)}, vol.~69, no.~10, pp.
  3940--3953, 2022.

\bibitem{chih202116}
Y.-D. Chih, P.-H. Lee, H.~Fujiwara, Y.-C. Shih, C.-F. Lee, R.~Naous, Y.-L.
  Chen, C.-P. Lo, C.-H. Lu, H.~Mori \emph{et~al.}, ``{An 89TOPS/W and
  16.3TOPS/mm$^2$ All-Digital SRAM-Based Full-Precision Compute-In Memory Macro
  in 22nm for Machine-Learning Edge Applications},'' in \emph{Proceedings of
  the International Solid-State Circuits Conference (ISSCC)}, 2021.

\bibitem{tu2022trancim}
F.~Tu, Z.~Wu, Y.~Wang, L.~Liang, L.~Liu, Y.~Ding, L.~Liu, S.~Wei, Y.~Xie, and
  S.~Yin, ``{TranCIM: Full-Digital Bitline-Transpose CIM-based Sparse
  Transformer Accelerator With Pipeline/Parallel Reconfigurable Modes},''
  \emph{Journal of Solid-State Circuits (JSSC)}, pp. 1--12, 2022.

\bibitem{jinshan2023isscc}
J.~Yue, C.~He, Z.~Wang, Z.~Cong, Y.~He, M.~Zhou, W.~Sun, X.~Li, C.~Dou,
  F.~Zhang \emph{et~al.}, ``{A 28nm 16.9-300TOPS/W Computing-in-Memory
  Processor Supporting Floating-Point NN Inference/Training with Intensive-CIM
  Sparse-Digital Architecture},'' in \emph{IEEE International Solid-State
  Circuits Conference (ISSCC)}, 2023, pp. 1--3.

\bibitem{shiweiisscc}
S.~Liu, P.~Li, J.~Zhang, Y.~Wang, H.~Zhu, W.~Jiang, S.~Tang, C.~Chen, Q.~Liu,
  and M.~Liu, ``{A 28nm 53.8TOPS/W 8b Sparse Transformer Accelerator with
  In-Memory Butterfly Zero Skipper for Unstructured-Pruned NN and CIM-Based
  Local-Attention-Reusable Engine},'' in \emph{IEEE International Solid-State
  Circuits Conference (ISSCC)}, 2023, pp. 250--252.

\bibitem{fengbin2023isscc}
F.~Tu, Z.~Wu, Y.~Wang, W.~Wu, L.~Liu, Y.~Hu, S.~Wei, and S.~Yin, ``{MulTCIM: A
  28nm 2.24$\mu$J/Token Attention-Token-Bit Hybrid Sparse Digital CIM-Based
  Accelerator for Multimodal Transformers},'' in \emph{IEEE International
  Solid-State Circuits Conference (ISSCC)}, 2023, pp. 248--250.

\bibitem{fengbin2023isscctensor}
F.~Tu, Y.~Wang, Z.~Wu, W.~Wu, L.~Liu, Y.~Hu, S.~Wei, and S.~Yin, ``{TensorCIM:
  A 28nm 3.7nJ/Gather and 8.3TFLOPS/W FP32 Digital-CIM Tensor Processor for
  MCM-CIM-Based Beyond NN Acceleration},'' in \emph{IEEE International
  Solid-State Circuits Conference (ISSCC)}, 2023, pp. 254--256.

\bibitem{shafaei2014fincacti}
A.~Shafaei, Y.~Wang, X.~Lin, and M.~Pedram, ``{FinCACTI: Architectural Analysis
  and Modeling of Caches with Deeply-scaled FinFET Devices},'' in
  \emph{Proceedings of the Computer Society Annual Symposium on VLSI (ISVLSI)},
  2014.

\bibitem{le202364}
M.~Le~Gallo, R.~Khaddam-Aljameh, M.~Stanisavljevic, A.~Vasilopoulos,
  B.~Kersting, M.~Dazzi, G.~Karunaratne, M.~Br{\"a}ndli, A.~Singh, S.~M.
  Mueller \emph{et~al.}, ``A 64-core mixed-signal in-memory compute chip based
  on phase-change memory for deep neural network inference,'' \emph{Nature
  Electronics}, pp. 1--14, 2023.

\bibitem{Garofalo2022AHI}
A.~Garofalo, G.~Ottavi, F.~Conti, G.~Karunaratne, I.~Boybat, L.~Benini, and
  D.~Rossi, ``A heterogeneous in-memory computing cluster for flexible
  end-to-end inference of real-world deep neural networks,'' \emph{IEEE Journal
  on Emerging and Selected Topics in Circuits and Systems (JETCAS)}, vol.~12,
  no.~2, pp. 422--435, 2022.

\bibitem{Hung2021AFC}
J.-M. Hung, C.-X. Xue, H.-Y. Kao, Y.-H. Huang, F.-C. Chang, S.-P. Huang, T.-W.
  Liu, C.-J. Jhang, C.-I. Su, W.-S. Khwa \emph{et~al.}, ``A four-megabit
  compute-in-memory macro with eight-bit precision based on cmos and resistive
  random-access memory for ai edge devices,'' \emph{Nature Electronics},
  vol.~4, no.~12, pp. 921--930, 2021.

\bibitem{deng2020model}
L.~Deng, G.~Li, S.~Han, L.~Shi, and Y.~Xie, ``Model compression and hardware
  acceleration for neural networks: A comprehensive survey,'' \emph{Proceedings
  of the IEEE}, vol. 108, no.~4, pp. 485--532, 2020.

\bibitem{yang2021s}
J.~Yang, W.~Fu, X.~Cheng, X.~Ye, P.~Dai, and W.~Zhao, ``{S2Engine: A novel
  systolic architecture for sparse convolutional neural networks},'' \emph{IEEE
  Transactions on Computers (TC)}, vol.~71, no.~6, pp. 1440--1452, 2021.

\bibitem{dai2020sparsetrain}
P.~Dai, J.~Yang, X.~Ye, X.~Cheng, J.~Luo, L.~Song, Y.~Chen, and W.~Zhao,
  ``{SparseTrain: Exploiting dataflow sparsity for efficient convolutional
  neural networks training},'' in \emph{Proceedings of 57th ACM/IEEE Design
  Automation Conference (DAC)}, 2020, pp. 1--6.

\bibitem{mishra2021accelerating}
A.~Mishra, J.~A. Latorre, J.~Pool, D.~Stosic, D.~Stosic, G.~Venkatesh, C.~Yu,
  and P.~Micikevicius, ``Accelerating sparse deep neural networks,''
  \emph{arXiv preprint arXiv:2104.08378}, 2021.

\end{thebibliography}

}


\vspace{-8mm}

\begin{IEEEbiography}[{\includegraphics[width=1in,height=1.25in,clip,keepaspectratio]{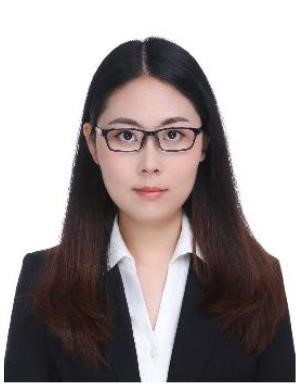}}]{Cenlin Duan}

received the B.S. degree in Electronic Science and Technology from University of Electronic Science and Technology of China, Chengdu, China, in 2015, and the M.S. degree in Software Engineering from Xidian University, Xi'an, China, in 2018. She is currently pursuing the Ph.D. degree at the School of Integrated Circuit Science and Engineering, Beihang University, Beijing, China. Her current research interests include processing-in-memory architectures and deep learning accelerators.

\end{IEEEbiography}

\vspace{-8mm}

\begin{IEEEbiography}[{\includegraphics[width=1in,height=1.25in,clip,keepaspectratio]{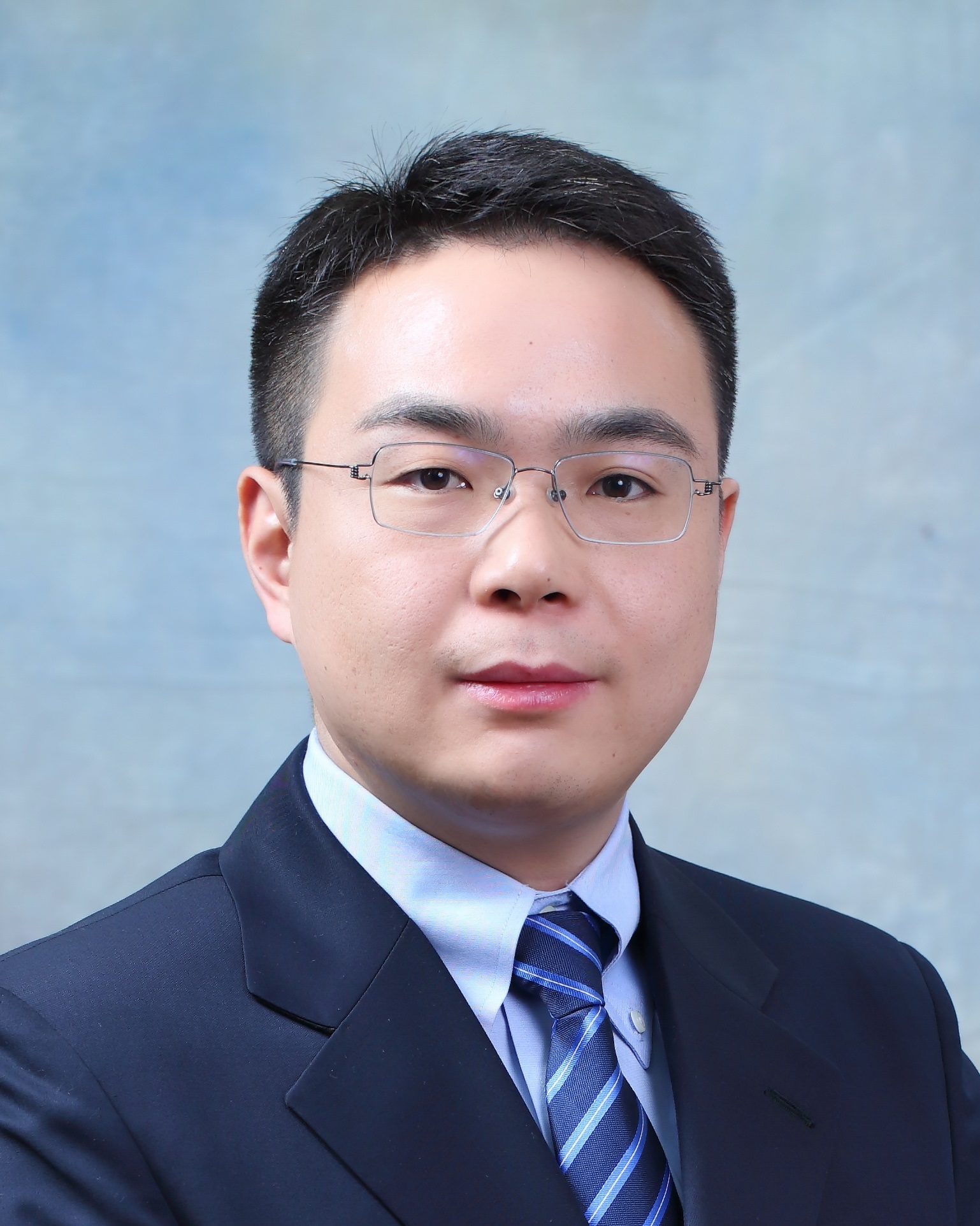}}]{Jianlei Yang}

(S'11-M'14-SM'20) received the B.S. degree in microelectronics from Xidian University, Xi'an, China, in 2009, and the Ph.D. degree in computer science and technology from Tsinghua University, Beijing, China, in 2014.

He is currently an Associate Professor in Beihang University, Beijing, China, with the School of Computer Science and Engineering. From 2014 to 2016, he was a post-doctoral researcher with the Department of ECE, University of Pittsburgh, Pennsylvania, USA.
His current research interests include deep learning accelerators and neuromorphic computing systems.

Dr. Yang was the recipient of the First/Second place on ACM TAU Power Grid Simulation Contest in 2011/2012. He was a recipient of IEEE ICCD Best Paper Award in 2013, ACM GLSVLSI Best Paper Nomination in 2015, IEEE ICESS Best Paper Award in 2017, ACM SIGKDD Best Student Paper Award in 2020.

\end{IEEEbiography}

\vspace{-8mm}

\begin{IEEEbiography}[{\includegraphics[width=1in,height=1.25in,clip,keepaspectratio]{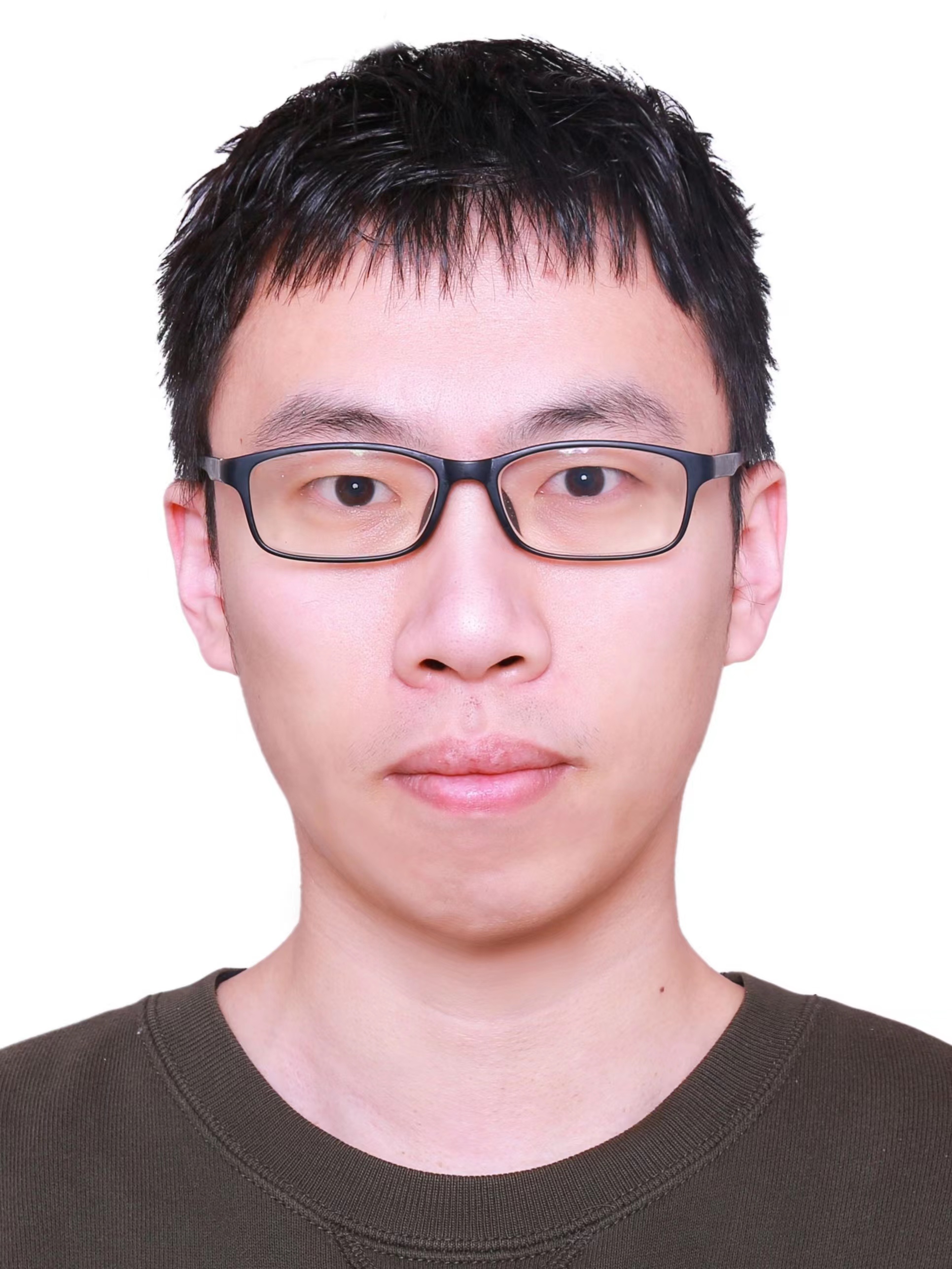}}]{Xiaolin He}

received the B.S. degree in software engineering from Beihang University, Beijing, China, in 2016. He is currently pursuing the Ph.D. degree at the School of Computer Science and Engineering, Beihang University, China. His research interests include in-memory computing architectures and compiler optimization techniques.

\end{IEEEbiography}

\vspace{-8mm}

\begin{IEEEbiography}[{\includegraphics[width=1in,height=1.25in,clip,keepaspectratio]{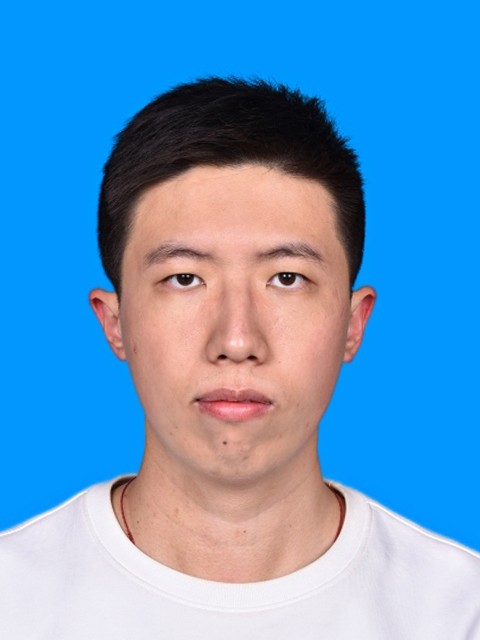}}]{Yingjie Qi}

received the B.S. degree in computer science and technology from Beihang University, Beijing, China, in 2020. He is currently pursuing the Ph.D. degree at the School of Computer Science and Engineering, Beihang University, China. His research interests include graph neural networks acceleration, processing-in-memory architectures and deep learning compilers.

\end{IEEEbiography}

\vspace{-8mm}

\begin{IEEEbiography}[{\includegraphics[width=1in,height=1.25in,clip,keepaspectratio]{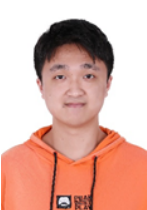}}]{Yikun Wang}

received the B.S. degree in computer science and technology from Beihang University, Beijing, China, in 2022. He is currently working toward the M.S. degree at the School of Computer Science and Engineering, Beihang University, China. His current research interests include computing-in-memory architectures and deep learning accelerators.

\end{IEEEbiography}

\vspace{-8mm}

\begin{IEEEbiography}[{\includegraphics[width=1in,height=1.25in,clip,keepaspectratio]{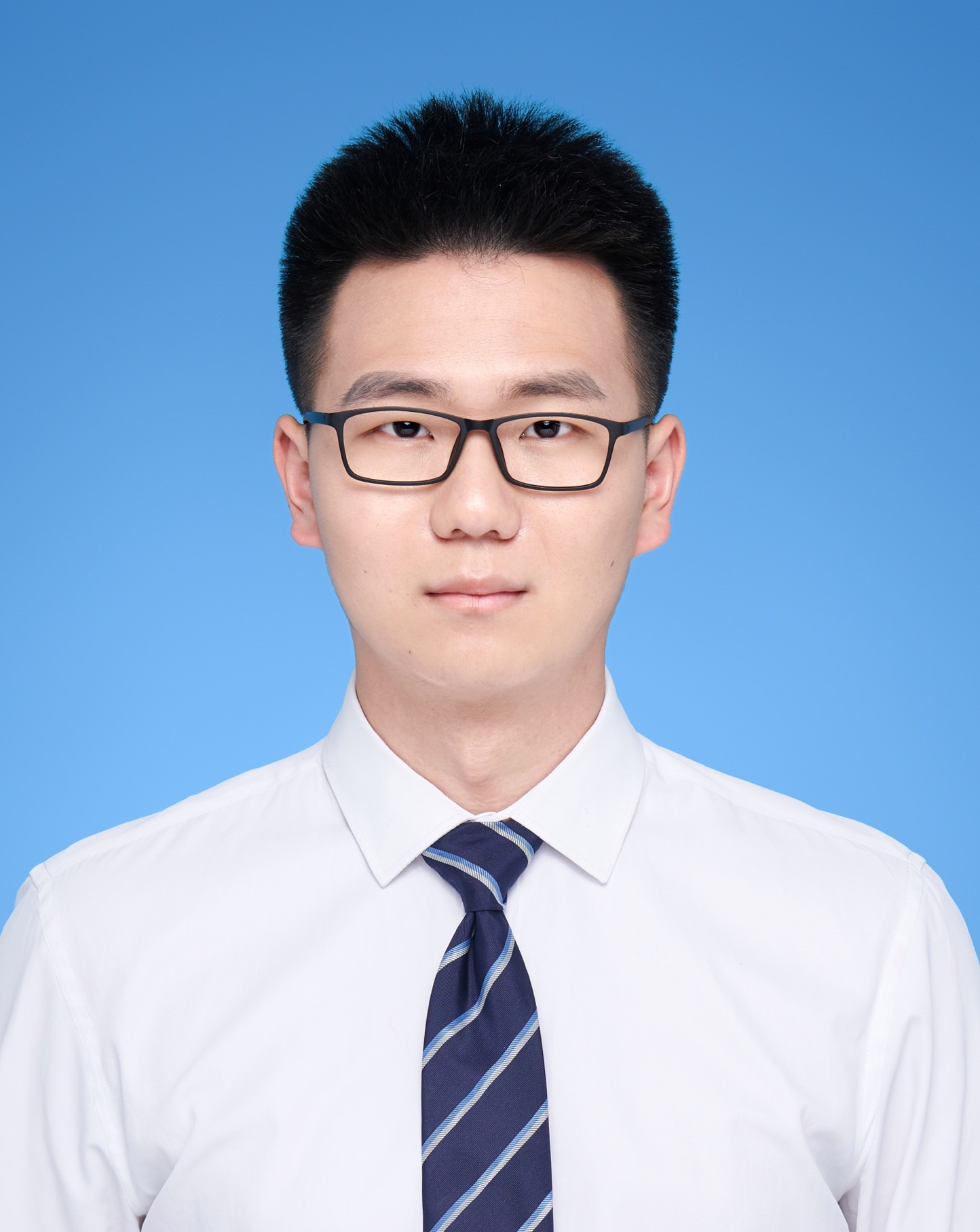}}]{Yiou Wang}

received the B.S. degree in computer science and technology from Beijing University of Technology, Beijing, China, in 2022. He is currently working toward the M.S. degree at the School of Computer Science and Engineering, Beihang University, China. His current research interests include deep learning compilers.

\end{IEEEbiography}

\vspace{-8mm}

\begin{IEEEbiography}[{\includegraphics[width=1in,height=1.25in,clip,keepaspectratio]{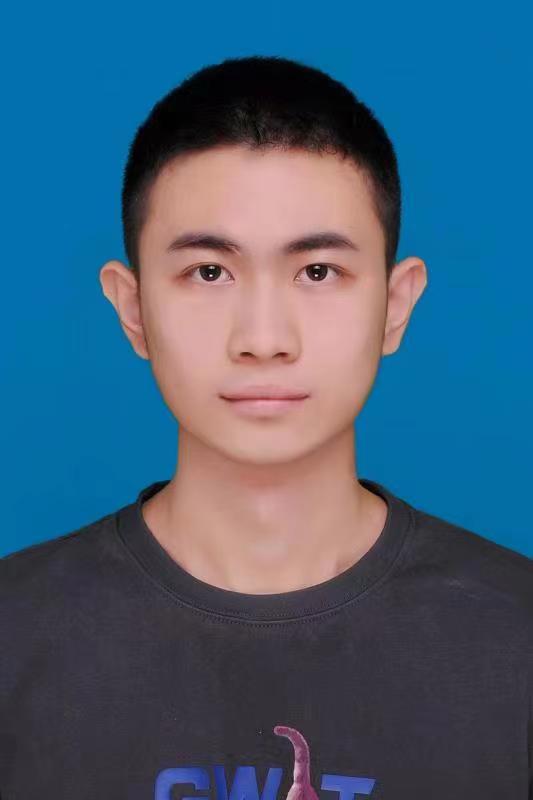}}]{Ziyan He}

received the B.S. degree in telecommunication engineering from Xidian University, Xi’an, China,in 2022. He is currently working toward the M.S. degree at the School of Telecommunication Engineering, Xidian University, Xi'an, China. His current research interests include processing-in-memory architecture and domain-specified accelerators.

\end{IEEEbiography}

\vspace{-8mm}

\begin{IEEEbiography}[{\includegraphics[width=1in,height=1.25in,clip,keepaspectratio]{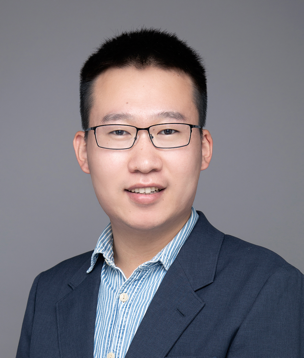}}]{Bonan Yan}

is currently an assistant professor at Institute for Artificial Intelligence, Peking University. He received his PhD degree from Department of Electrical and Computer Engineering, Duke University in 2020. Hi research interests include circuits and systems for artificial intelligence chips, VLSI design for emerging memory, especially processing-in-memory technology.

He has published more than 40 papers in renowned academic journals and conferences, including ISSCC, Symposium on VLSI Technology, IEDM, DAC, etc. He actively serves as a TPC member for the conferences, including DAC, AICAS, and EDTM.

\end{IEEEbiography}

\vspace{-8mm}

\begin{IEEEbiography}[{\includegraphics[width=1in,height=1.25in,clip,keepaspectratio]{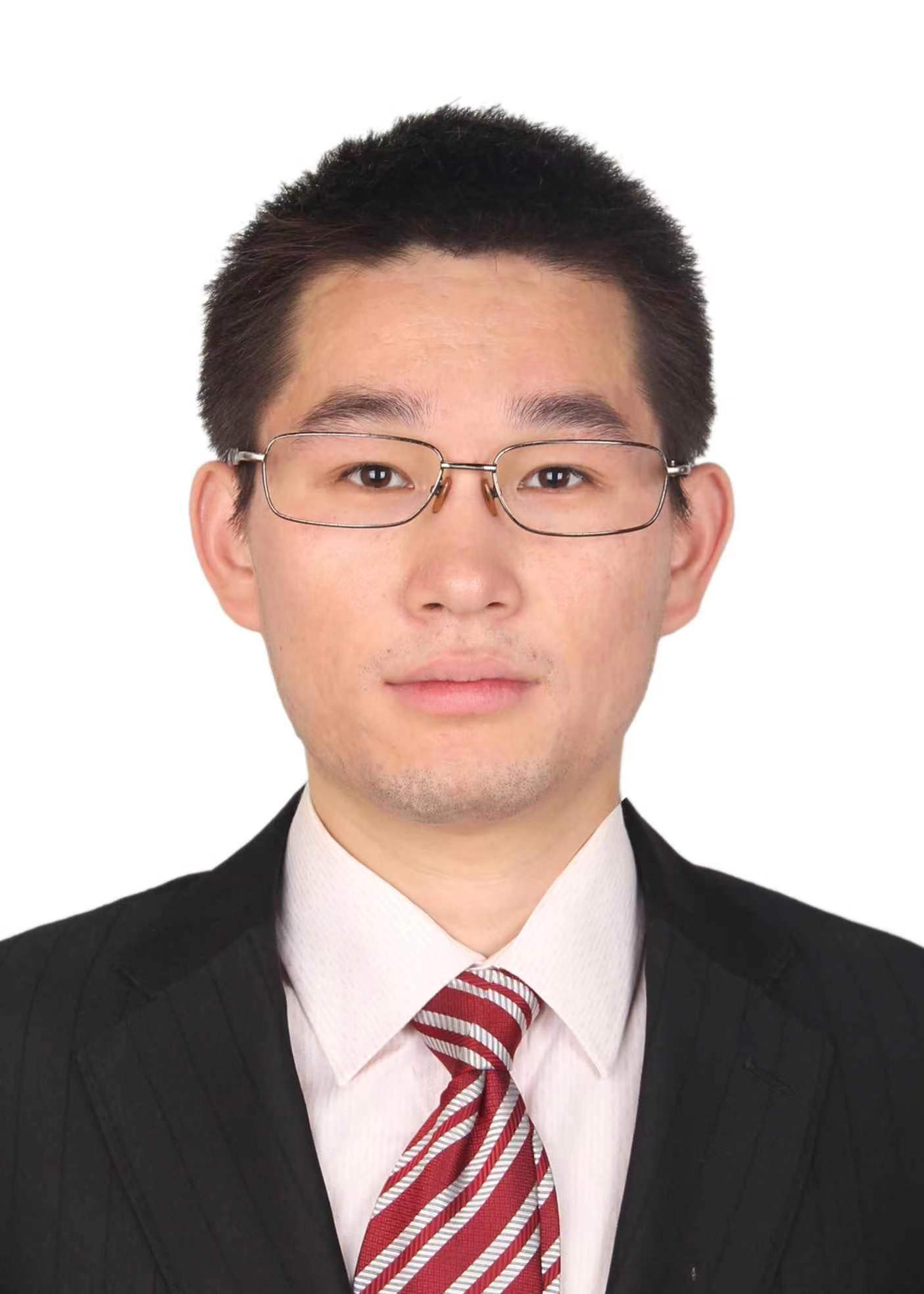}}]{Xiaotao Jia}

(S'13-M'17) received the B.S. degree in mathematics from Beijing Jiao Tong University, Beijing, China, in 2011, and the Ph.D. degree in computer science and technology from Tsinghua University, Beijing, China, in 2016. He is currently an associate professor in Beihang University, Beijing, China. His current research interests include spintronic circuits and Bayesian learning systems.

\end{IEEEbiography}

\vspace{-8mm}

\begin{IEEEbiography}[{\includegraphics[width=1in,height=1.25in,clip,keepaspectratio]{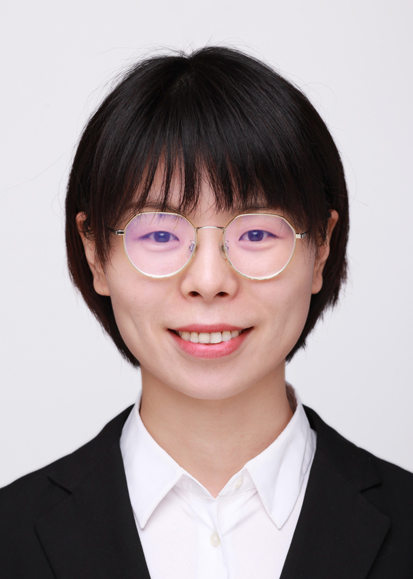}}]{Xueyan Wang}

received the B.S. degree in computer science and technology from Shandong University, Jinan, China,in 2013, and the Ph.D. degree in computer science and technology from Tsinghua University, Beijing, China, in 2018. From 2015 to 2016, she was a visiting scholar in University of Maryland, College Park, MD, USA.

She is currently an Assistant Professor with the School of Integrated Circuit Science and Engineering in Beihang University, Beijing, China. Her current research interests include processing-in-memory architectures and hardware security.

\end{IEEEbiography}

\vspace{-8mm}

\begin{IEEEbiography}[{\includegraphics[width=1in,height=1.25in,clip,keepaspectratio]{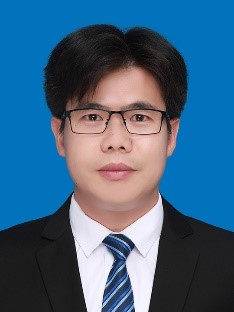}}]{Weitao Pan}

received the B.S. degree from School of Technical Physics of Xidian University in 2004. His Ph.D. degree was received from School of Microelectronics of Xidian University in 2010. Now he is an associate professor in State Key Laboratory of Integrated Service Networks of Xidian University. His current research interests include VLSI design methods and post-silicon verification.

\end{IEEEbiography}

\vspace{-8mm}

\begin{IEEEbiography}[{\includegraphics[width=1in,height=1.25in,clip,keepaspectratio]{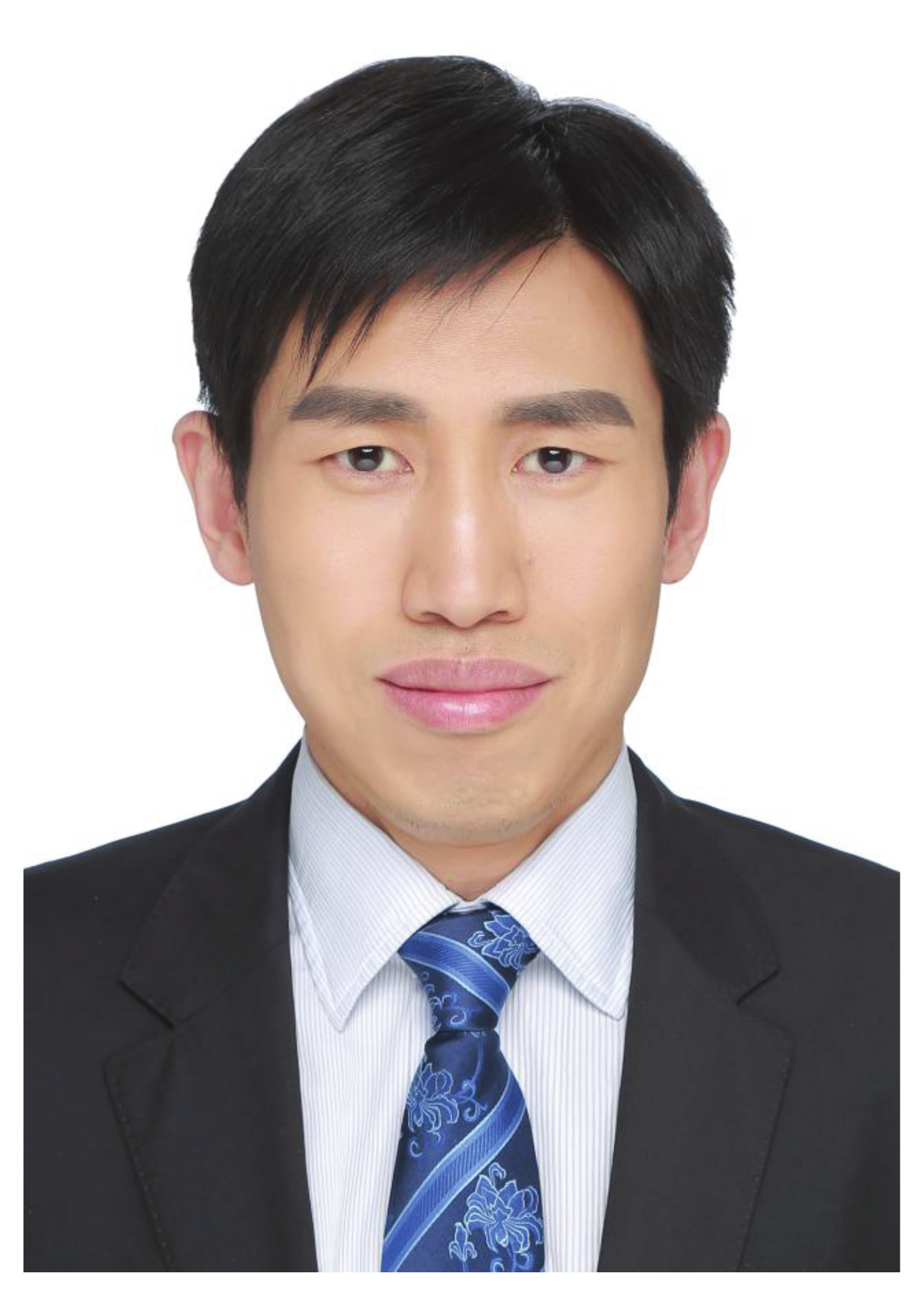}}]{Weisheng Zhao}

(Fellow, IEEE) received the Ph.D. degree in physics from the University of Paris Sud, Paris, France, in 2007.

He is currently a Professor with the School of Integrated Circuit Science and Engineering, Beihang University, Beijing, China. In 2009, he joined the French National Research Center, Paris, as a Tenured Research Scientist. Since 2014, he has been a Distinguished Professor with Beihang University. He has published more than 200 scientific articles in leading journals and conferences, such as \textit{Nature
Electronics}, \textit{Nature Communications}, \textit{Advanced Materials}, IEEE Transactions, ISCA, and DAC. His current research interests include the hybrid integration of nanodevices with CMOS circuit and new nonvolatile memory (40-nm technology node and below) like MRAM circuit and architecture design.

Prof. Zhao is currently the Editor-in-Chief for the {\sc{IEEE Transactions on Circuits and System I: Regular Paper}}.

\end{IEEEbiography}

\end{document}